\documentclass[journal=jacsat,manuscript=communication]{achemso}
\SectionsOn
\usepackage{sectsty}
\sectionfont{\sffamily\large}
\subsectionfont{\sffamily\normalsize}
\usepackage[version=3]{mhchem} % Formula subscripts using \ce{}

\usepackage[separate-uncertainty=true]{siunitx}

\usepackage{chemfig}

  \usepackage[%  
     colorlinks=true,
     pdfborder={0 0 0},
     linkcolor=blue,
     citecolor=green
  ]{hyperref}
  \setkeys{acs}{doi = true} %% achemso uses natbib acs style
  \newcommand*{\doi}[1]{\href{https://dx.doi.org/#1}{#1}}
\author{Zachery A. Enderson}
\author{Harshavardhan Murali}
\affiliation[Georgia Tech]
{School of Physics, Georgia Institute of Technology, Atlanta, GA  30332}
\author{Raghunath R. Dasari}
\affiliation[Georgia Tech]
{School of Chemistry and Biochemistry, Georgia Institute of Technology, Atlanta, GA  30332}
\author{Qingqing Dai}
\author{Hong Li}
\affiliation[University of Arizona]
{Department of Chemistry and Biochemistry, The University of Arizona, Tucson, AZ 85721}
\author{Timothy C. Parker}
\affiliation[Georgia Tech]
{School of Chemistry and Biochemistry, Georgia Institute of Technology, Atlanta, GA  30332}
\author{Jean-Luc Br\'{e}das}
\affiliation[University of Arizona]
{Department of Chemistry and Biochemistry, The University of Arizona, Tucson, AZ 85721}
\author{Seth R. Marder}
\affiliation[Georgia Tech]
{School of Chemistry and Biochemistry, Georgia Institute of Technology, Atlanta, GA  30332}
\alsoaffiliation[CU]
{University of Colorado Boulder, Renewable and Sustainable Energy Institute, Department of Chemical and Biological Engineering, Department of Chemistry, and Materials Science and Engineering Program,  Boulder CO 80303}
\alsoaffiliation[NREL]
{National Renewable Energy Laboratory, Chemistry and Nanoscience Center, Golden CO 80401}
\author{Phillip N. First}
\affiliation[Georgia Tech]
{School of Physics, Georgia Institute of Technology, Atlanta, GA  30332}
\email{first@gatech.edu}

\title[]
{
    Tailoring On-Surface Molecular Reactions and Assembly through Hydrogen-Modified Synthesis: From Triarylamine Monomer to {2D} Covalent Organic Framework
}

\abbreviations{COF, DTPA, SAM, HV, UHV, STM}

\begin{document}

%%% START arxiv: Locate TOC graphic after bibliography (by hand)
% \begin{tocentry}
% %\includegraphics[height = 3.5 cm]{"figures/Graphical TOC V2D.png"}
% \includegraphics[height = 3.5 cm]{Figure_TOC}
% \end{tocentry}
%%% END arxiv

\begin{abstract}
  Relative to conventional wet-chemical synthesis techniques, on-surface synthesis of organic networks in ultrahigh vacuum has few control parameters. The molecular deposition rate and substrate temperature are typically the only synthesis variables to be adjusted dynamically. Here we demonstrate that reducing conditions in the vacuum environment can be created and controlled without dedicated sources---relying only on backfilled hydrogen gas and ion gauge filaments---and can dramatically influence the Ullmann-like on-surface reaction used for synthesizing two-dimensional covalent organic frameworks (2D COFs). Using tribromo dimethylmethylene-bridged triphenylamine (\ce{(Br_3)DTPA}) as monomer precursors, we find that atomic hydrogen (\ce{\Lewis{0.,H}}) blocks aryl-aryl bond formation to such an extent that we suspect this reaction may be a factor in limiting the ultimate size of 2D COFs created through on-surface synthesis. Conversely, we show that control of the relative monomer and hydrogen fluxes can be used to produce large self-assembled islands of monomers, dimers, or macrocycle hexamers, which are of interest in their own right. On-surface synthesis of oligomers, from a single precursor, circumvents potential challenges with their protracted wet-chemical synthesis and with multiple deposition sources. Using scanning tunneling microscopy and spectroscopy (STM/STS), we show that changes in the electronic states through this oligomer sequence provide an insightful view of the 2D-COF (synthesized in the absence of atomic hydrogen) as the endpoint in an evolution of electronic structures from the monomer.
\end{abstract}

\section{Introduction}
\label{sec:Intro}
The targeted synthesis of materials with designed properties or electronic structure is a central goal of materials research. For ordered covalent organic framework (COF) materials, limiting the chemical synthesis to two dimensions (2D) is especially challenging, but its mastery would advance technologies for 2D organic electronics \cite{Ding2013,YangF2018,Wang2021} and for membranes used both structurally and for chemical separations \cite{Yuan2019}. One approach to fabricating low-dimensional COFs is on-surface synthesis: a version of reticular synthesis \cite{Yaghi2003} conducted \latin{in vacuo} on a clean surface. Rigid molecular precursors are vapor-deposited onto a substrate, generally metallic, which facilitates bond formation between molecular units \cite{Shen2017}. This method allows for COFs to be carefully designed with specific lattice structures, functional groups, and electronic structure \cite{Thomas19a}. In this work, we focus on heterotriangulene 2D polymers from tribromo-substituted dimethylmethylene-bridged triphenylamine (\ce{(Br_3)DTPA}) monomer precursors, first studied by Bieri \latin{et al.} \cite{Bieri2011}. Interest in heterotriangulene COFs arises from their potential optoelectronic and transport properties, which are chemically-tunable through their electronic structure: The bandgap, and the occurrence of both flat bands (with potential to host correlated states) and linearly dispersive (Dirac) bands, is controlled by modifications of the bridge-site moieties or the central heteroatom site \cite{Field2002,Kan2012,Hammer2015,Steiner2017,Pavlicek2017,Jing2019,Hirai2019,Schaub2020}. In addition, for a bridge-moiety resulting in an open-shell configuration, the 2D-COF is predicted to have a fully spin-polarized bandstructure near the Fermi energy, \(E_F\)\cite{Kan2012}.

While on-surface synthesis is a valuable approach to create low-dimensional COFs, it is not without limitations. The formation of effectively irreversible covalent bonds makes the extent and the order of 2D COFs highly dependent on the deposition parameters and other environmental factors \cite{Bjork2013,Wang2017,Cai2017,Clair2019}. In particular, it has been suggested \cite{Simonov2018} and shown by scanning probe microscopy \cite{Talirz2013,Kong2017, Kawai2017,Zuzak2020} that the presence of atomic hydrogen during deposition can inhibit \ce{C-C} bond formation. Here we use time-of-flight secondary-ion mass spectroscopy (TOF-SIMS) to provide conclusive evidence for the nature of atomic hydrogen's inhibitory effect on COF growth. We also show that the influence of atomic hydrogen during the formation of 2D-COFs can be pervasive: Hot filaments \emph{anywhere} within a vacuum chamber can potentially crack \ce{H_2}. Then, from a broader perspective, we demonstrate that atomic hydrogen, even with ambient cracking filaments, could become a valuable reducing agent for control of surface chemistry, enabling the on-surface creation of single- and multi-monomer compounds from a single precursor source. Our scanning tunneling microscopy/spectroscopy (STM/STS) investigation of the evolution of electronic structures through such a series of oligomers (monomer, dimer, hexamer macrocycle) provides a deeper understanding of the 2D COF created from the same molecular precursor.

\section{Results and Discussion}

\subsection{Hydrogen Effect}
\label{sec:Hydrogen}

Prior work \cite{Bieri2011,Schlutter2013,Steiner2017,Wang2018,Lackinger2017,Fritton2019} has established a general understanding of the on-surface reactions for halogen-terminated heterotriangulene monomer precursors. Above \SI{330}{\kelvin} the molecules dehalogenate through an Ullmann-like reaction with the Ag(111) substrate, temporarily leaving them as surface-stabilized radicals \cite{Wang2018}. On Ag(111), surface diffusion enables these species to form organometallic networks with monomer--Ag--monomer linkages. Some examples include the honeycomb organometallic lattice of DTPA (formed after depositing precursors at \SI{473}{\kelvin}) \cite{Bieri2011} and 1,3,5-tris(4-bromophenyl)benzene (\SI{375}{\kelvin}) \cite{Fritton2019}. When heated to higher temperatures (\SI{500}{\kelvin} or higher as suggested by XPS studies \cite{Fritton2019}), the molecules have enough energy to complete the Ullmann coupling, forming \ce{C-C} bonds at their debrominated sites and creating a honeycomb-lattice 2D-COF \cite{Bieri2011}. 

\begin{figure}
  \includegraphics[width=0.85\columnwidth]{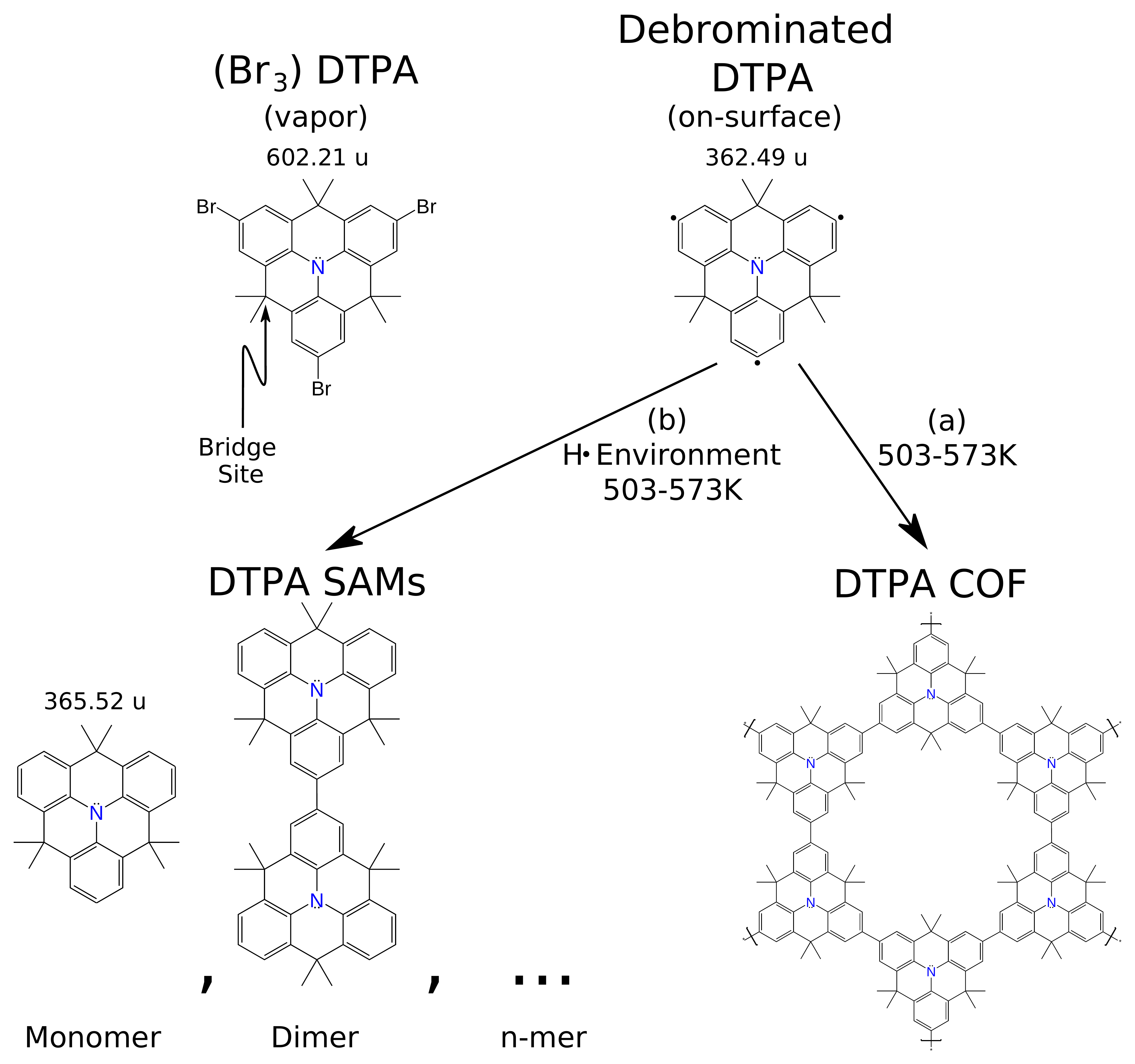}
  \caption{On-surface reaction of tribromo-DTPA precursor molecules on
    a heated metallic substrate in an atomic hydrogen rich or high
    vacuum environment. \ce{(Br_3)DTPA} molecules are evaporated onto
    a Ag(111) substrate held at a constant temperature within
    \SIrange{503}{573}{\kelvin}. First reaction step: Dehalogenation
    of \ce{(Br_3)DTPA} resulting in adsorbed DTPA radicals. Second
    reaction step: Bonding at debrominated sites, dependent on the
    environmental conditions: (a) completion of Ullmann-type coupling
    forming \ce{C-C} bonds between DTPA molecules or (b) reaction with
    atomic hydrogen forming \ce{C-H} bonds on a fraction of the
    radical sites. This terminates the polymer at $n$ units, with $n$
    determined by kinetics.}
\label{fig:COFsynth}
\end{figure}

The monomer precursor used in this work and its proposed on-surface radical configuration are shown in Figure~\ref{fig:COFsynth}, as well as the observed products for (a) UHV deposition and (b) deposition in the presence of atomic hydrogen. In these experiments, \ce{(Br_3)DTPA} precursor molecules were deposited from heated crystallized monomers onto temperature-controlled Ag(111) substrates. When deposited on surface, the methyl groups point perpendicular to the surface and dominate the STM imaging, forming the triangular shapes for the molecules. An important point is that experiments were done in a multi-user STM facility; therefore, to limit potential contamination of the main UHV chambers, depositions were conducted in the turbo-pumped sample-introduction chamber (load-lock) with an ambient pressure of \SIrange[range-units=single,range-phrase=--]{2}{3e-8}{\milli\bar} (HV).  These conditions produced results which are cautionary in a sense, yet promise an additional level of control over on-surface synthesis.

\begin{figure*}
  \includegraphics[width=0.85\textwidth]{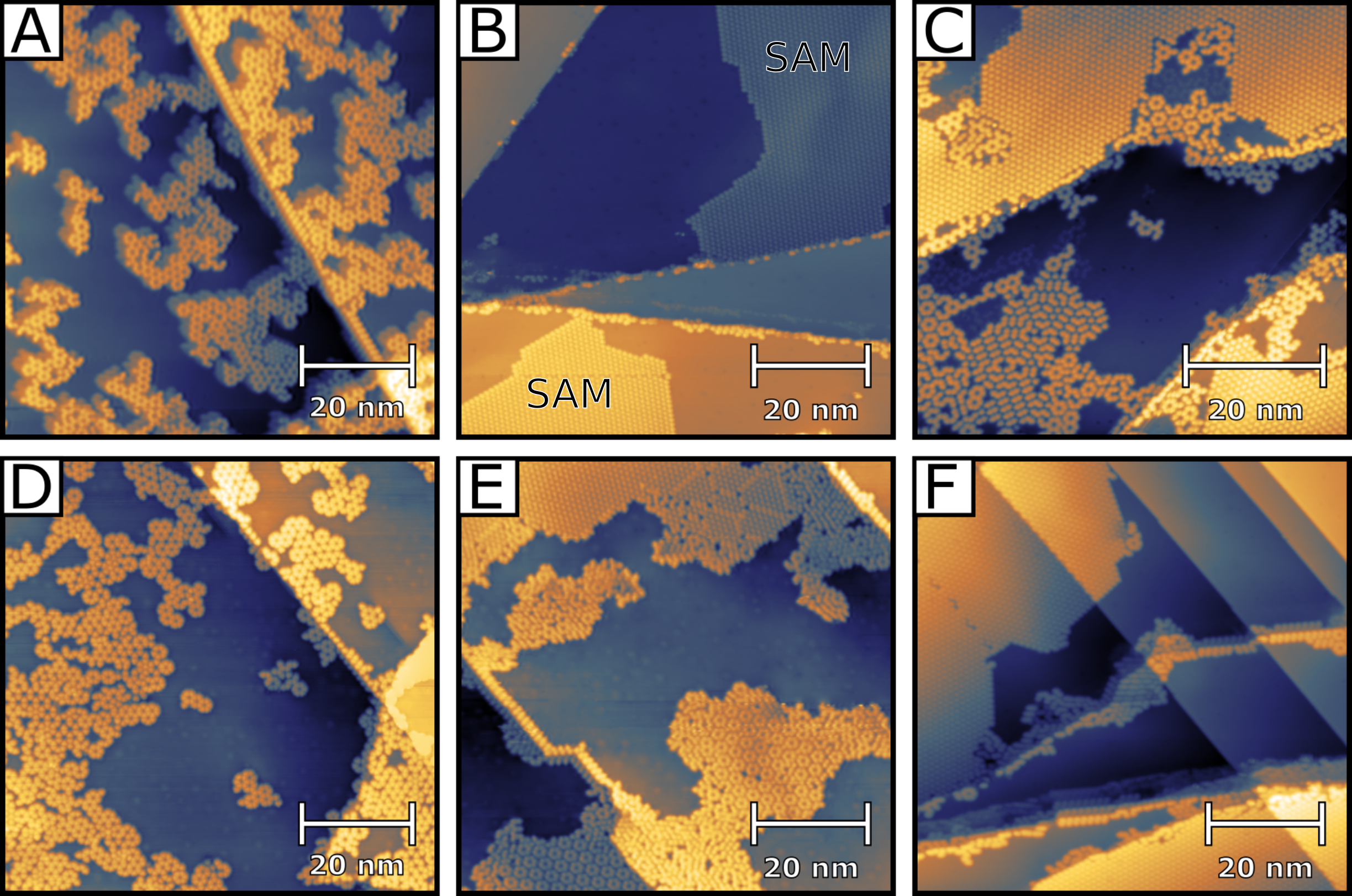}
  \caption{STM topographs of DTPA depositions under various
    environmental conditions. All depositions are on a Ag(111)
    substrate at temperatures \SI{523}{\kelvin}(A-C) and
    \SI{503}{\kelvin}(D-F) with a deposition rate of 0.055 ML/min
    (A-C) and 0.025 ML/min (D-F). (A-C) Depositions in a chamber
    backfilled with \ce{H_2} to \SI{2.2e-6}{\milli\bar} with (A) the
    cracking filament off throughout the process, (B) the cracking
    filament on (\SI{0.1}{\milli\ampere} emission current) during
    \ce{DTPA} deposition, and (C) the clean Ag(111) sample in the
    chamber with the cracking filament on (\SI{0.1}{\milli\ampere}
    emission current) for 10 minutes prior to a \ce{DTPA} deposition
    with the filament off. (D-F) Depositions in a HV chamber
    ($P = \SI{2.9e-8}{\milli\bar}$) with (D) the cracking filament off
    throughout the process, (E) the cracking filament on with
    \SI{0.1}{\milli\ampere} emission current, and (F) the cracking
    filament on with \SI{1}{\milli\ampere} emission current (for more
    information, see Methods).}
  \label{fig:igOnOff}
\end{figure*}

Initial deposition attempts showed that the (\ce{(Br_3)DTPA}) covalent bonding process on Ag(111) was inhibited, a result that was traced to the on/off state of a remote (no line-of-sight path to the silver substrate) thoriated-iridium filament in the load-lock ion gauge (see Supporting Information Figure~S1). Adopting the hypothesis that in HV this filament acts as a ``cracking'' source (\ce{H_2 -> 2 \Lewis{0.,H}}), inhibiting aryl-aryl bonding through competitive \ce{C-H} bonds, we conducted the experiments summarized in Figure~\ref{fig:igOnOff}. These depositions were conducted with varying amounts of environmental atomic hydrogen, controlled by changing the amount of molecular hydrogen in the chamber or the filament temperature (by increasing the emission current setting of the ion gauge). During deposition, the substrate was held at a constant temperature, chosen within the lower end of the range indicated in Figure~\ref{fig:COFsynth}, \SIrange{503}{523}{\kelvin}, because the hydrogen terminated \ce{DTPA} monomers were found to desorb from the surface at higher temperatures (see Supporting Information Figure~S2). The primary finding is seen by the different concentrations of DTPA structures present between Figure~\ref{fig:igOnOff}A and Figure~\ref{fig:igOnOff}B. Figure~\ref{fig:igOnOff}A is the control deposition with the cracking filament off throughout the time the sample is in the deposition chamber (before, during, and after deposition) and in the presence of additional \ce{H_2} backfilled to a pressure of \SI{2.2e-6}{\milli\bar}. The observed \SIrange{10}{20}{\nano\metre} COF islands are consistent with prior results \cite{Bieri2011,Steiner2017} and no other significant structures are observed. In contrast to this, Figure~\ref{fig:igOnOff}B shows a deposition conducted in the same elevated molecular hydrogen environment and the cracking filament turned on only during \ce{DTPA} deposition. In this case, only large islands of monomer self-assembled monolayers (SAMs) are observed. An additional experiment presented in Figure~\ref{fig:igOnOff}C found an unexpected result. The image in Figure~\ref{fig:igOnOff}C was acquired after pre-exposing the sample to \SI{2.2e-6}{\milli\bar} \ce{H_2} gas with the cracking filament on for \SI{10}{\minute} prior to deposition, turning off the filament, pumping the \ce{H_2} from the chamber, and then performing the \ce{DTPA} deposition. This resulted in a mixture of structures showing more oligomer SAM formations than the control deposition in Figure~\ref{fig:igOnOff}A but also more covalent coupling than Figure~\ref{fig:igOnOff}B.

The results shown in the top row of Figure~\ref{fig:igOnOff} support the hypothesis that it is atomic hydrogen, produced at the remote cracking filament, which inhibits on-surface \ce{C-C} bond formation and hence COF formation. We confirmed that the source of atomic hydrogen is thermal cracking of \ce{H2} at the hot filament, rather than energetic electrons or ions produced by accelerating potentials in the ion gauge (see Figure~S3).  Due to its lower operating temperature (\(\SI{\sim 1700}{K}\) at \SI{0.1}{mA} emission), thoriated iridium has a lower cracking efficiency than tungsten operating at the same emission \cite{Sutoh1995}, yet clearly produces sufficient \ce{\Lewis{0.,H}} to dramatically influence on-surface reactions. Furthermore, the results from Figure~\ref{fig:igOnOff}C show that the Ag(111) surface (or possibly other nearby surfaces) serves as a reservoir for \ce{\Lewis{0.,H}}.  Once filled by pre-exposure to a source of atomic hydrogen, this reservoir allows \ce{\Lewis{0.,H}} to affect the on-surface reaction long after the cracking filament is off and the \ce{H_2} has been evacuated. 

Other depositions, in HV conditions and with the cracking filament on, show that the fractions of COF and SAMs (monomer, dimer, hexamer) which form depend sensitively on the deposition rate, the substrate temperature (Figure~S2), and the \ce{\Lewis{0.,H}} partial pressure. An example of this is seen by the controlled depositions shown in Figures~\ref{fig:igOnOff}D--\ref{fig:igOnOff}F. Each deposition uses the same deposition rate, substrate temperature, and background \ce{H2} pressure but varies the temperature of the cracking filament by changing its emission current (see Methods). As the filament temperature increases, its \ce{H_2} cracking efficiency increases, resulting in higher \ce{\Lewis{0.,H}} production. The increase in \ce{\Lewis{0.,H}} causes greater disruption in the DTPA covalent coupling which is seen by the progression from COF islands in Figure~\ref{fig:igOnOff}D, to a mixture of oligomer SAMs in Figure~\ref{fig:igOnOff}E, and finally to predominantly monomer SAMs in Figure~\ref{fig:igOnOff}F (See also Figures~S4--S7).  This is consistent with two chemical species (debrominated DTPA and atomic \ce{H}) competing to bond to the same radical sites. However, quantitative confirmation for any of the reaction products proposed in Figure~\ref{fig:COFsynth} cannot be determined from STM alone.
\begin{figure}
  \includegraphics{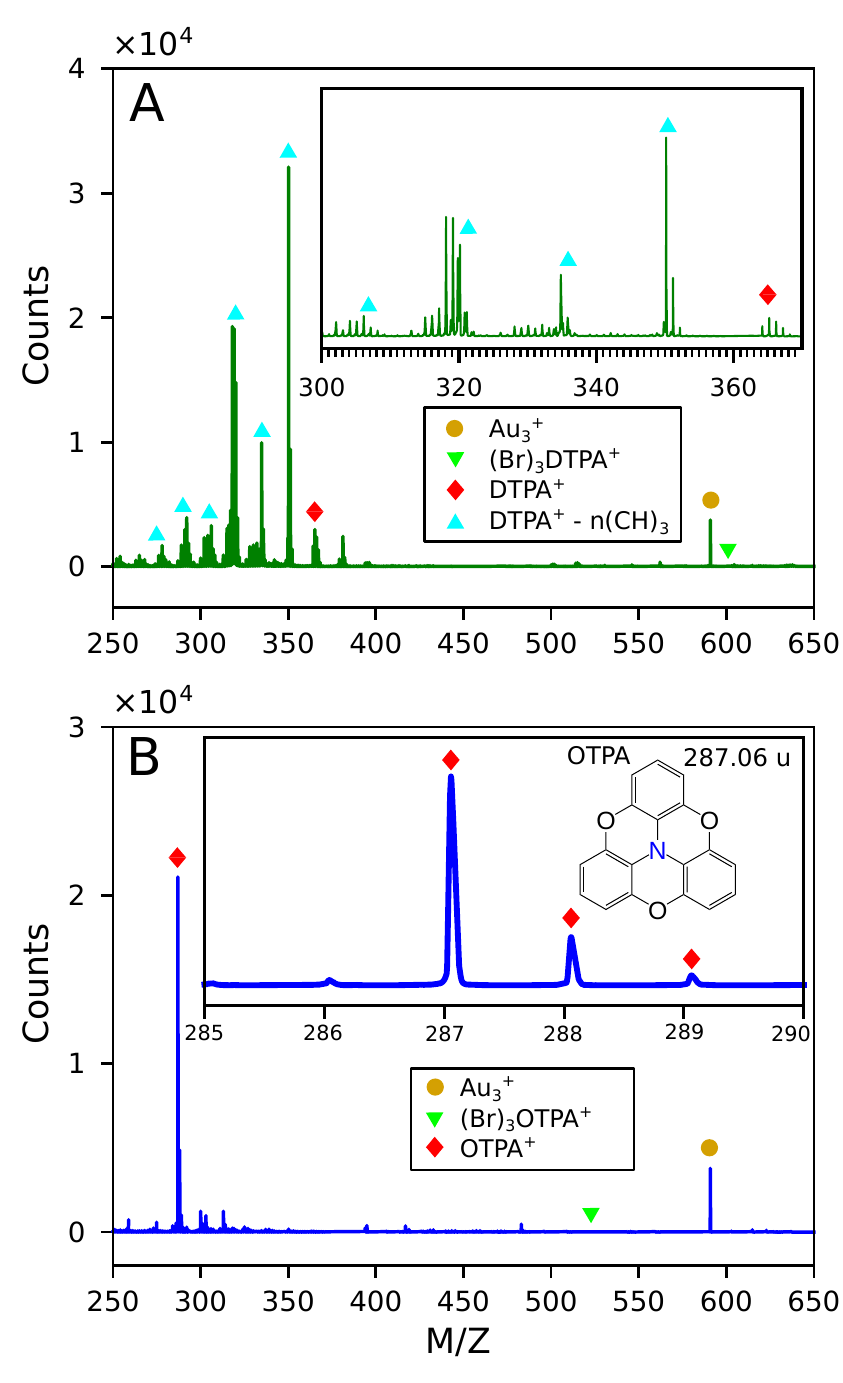}
  \caption{Positive ion TOF-SIMS analysis of a monolayer of (A) DTPA
    monomer SAM and (B) OTPA monomer SAM. The layer was obtained by
    the deposition of the brominated precursors on a heated Au(111)
    surface in the presence of a cracking filament in a \ce{H2}
    back-filled environment (\SI{2.2e-6}{\milli\bar}). The DTPA SAM
    shows fragmentation characterized by groups of peaks separated by
    roughly \SI{15}{\amu} which suggests the cleavage of bridging
    methyl groups due to the energetic secondary ion generation
    process. In contrast, the OTPA spectrum shows only one prominent
    set of peaks, due to the absence of weakly bound bridging methyl
    groups, which corresponds to \ce{OTPA^+} and its side peaks due to
    natural isotopic mass distribution. The absence of the brominated
    monomers in both A and B confirms the replacement of bromine with
    hydrogen.}
  \label{fig:TOFSIMS}
\end{figure}

To confidently identify the on-surface reaction products, we turn to TOF-SIMS.  Monomer DTPA samples similar to the one shown in Figure~\ref{fig:igOnOff}B were prepared on Au(111), imaged by STM (Figure~S8), and then transferred through atmosphere to the TOF-SIMS instrument. The samples were prepared on Au(111) instead of Ag(111) to avoid surface contamination by oxygen. Figure~\ref{fig:TOFSIMS}A displays the relevant portion of the mass spectrum, with the inset expanding the mass scale near the monomer \ce{DTPA^+} (\latin{cf.} Figure~\ref{fig:COFsynth}).  A key observation is vanishingly small intensities at masses corresponding to \ce{(Br_3)DTPA^+} or any other partially-brominated DTPA species. For masses smaller than monomer-\ce{DTPA^+}, the spectrum shows fragmentation due to the cleavage of increasing numbers of bridging methyl groups, resulting in evenly spaced peak sets, as denoted by light-blue triangles in Figure~\ref{fig:TOFSIMS}A. Within the peak set encompassing the exact mass (red diamonds), all peaks are well-described by natural isotopic abundances with the largest peak at 365.5u corresponding to the fully hydrogenated \ce{DTPA^+}. However, the more intense peak sets at lower masses are further complicated by metastable states, which have intensity at fractional masses due to in-flight transformations\cite{Shard2008}. Despite these complications, the mass spectrum of prepared monomer SAMs indicates that the monomers are, as hypothesized, simply hydrogen terminated DTPA molecules. That is, \ce{H} has replaced \ce{Br} in the \ce{(Br_3)DTPA} precursor molecules (see also Figure~S9).

Still, the complexity of the mass spectrum in Figure~\ref{fig:TOFSIMS}A may leave some doubt as to the robustness of this conclusion. To provide further support, we performed an analogous experiment on monomer SAMs of the related oxygen-bridged molecule \ce{OTPA} \cite{Galeotti2020}, shown in the inset to Figure~\ref{fig:TOFSIMS}B (precursors \ce{(Br_3)OTPA}, deposition conditions similar to those of Figure~\ref{fig:igOnOff}B on Au(111); see also Figure~S10). Without the bridging methyl groups, the cracking pattern seen in mass spectra from \ce{OTPA} samples (Figure~\ref{fig:TOFSIMS}B) is far simpler.  A single prominent hydrogen terminated \ce{OTPA^+} mass peak is found at the expected mass of 287.06u, with associated minor peaks due to the natural isotopic abundances of the constituent atoms. Also, like mass spectra for \ce{DTPA}, there is virtually zero intensity corresponding to any brominated species. This provides further support that \ce{H} has replaced \ce{Br} in the monomer product. Repeating the experiment using deuterium instead of hydrogen as a background gas shows the presence of OTPA terminated with deuterium instead of hydrogen, providing further confirmation of the proposed process (see Figure~S11 and accompanying discussion). The results of the deuterium experiment also confirm that the hydrogenation (and deuteration) occurs inside the vacuum chamber and not during transfer of the sample to the TOF-SIMS instrument ($\sim 5$ minutes exposure to atmosphere).

\subsection{Controlling On-surface Synthesis}
\label{sec:ControlSynth}

\begin{figure*}
  \includegraphics[width=0.85\textwidth]{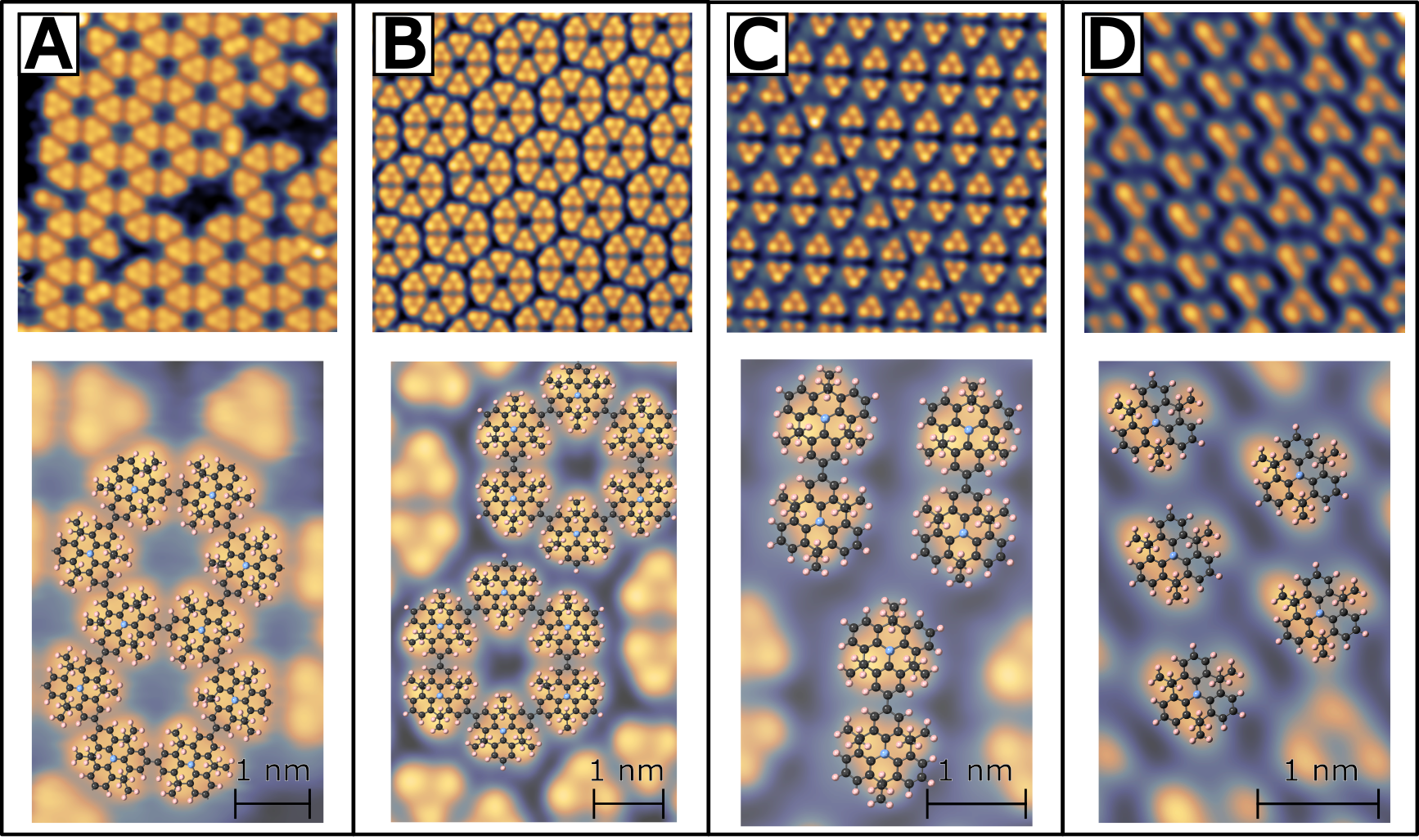}
  \caption{Experimental STM topographs of DTPA SAMs and COF
    superimposed with models. (A) DTPA COF (B) Hexamer SAM (C)
    Dimer/Trimer SAM (D) Monomer SAM. First row: Experimental
    topographs; Second row: Enlarged topograph superimposed with
    structural model (hydrogen-terminated DTPA for SAMs). The DTPA SAM
    images were acquired at 77K, while the COF data was acquired at
    \SI{4.7}{\kelvin}. (For specific deposition parameters, see
    Methods.)}
  \label{fig:SAMmodels}
\end{figure*}

As established above, the presence of atomic hydrogen during on-surface synthesis inhibits the Ullmann reaction between organic molecules. Now we show that, by controlling deposition parameters, this effect can be used to synthesize a useful sequence of oligomers from the same \ce{(Br_3)DTPA} precursors. Figure~\ref{fig:SAMmodels} shows the variety of compounds and their SAMs formed by deposition in controlled reducing environments. These topographs were chosen to emphasize the characteristic molecular structures and their assembly, not the entire surface distribution of oligomers/COF, which depends on the deposition parameters and is statistical in nature---as shown in Figure~\ref{fig:igOnOff} and further quantified below. 

Clearly, the molecular lattices of the SAMs vary for different oligomers, yet they share some similarities. Our first observation is simply the formation of SAMs, which implies that the intermolecular attractions are sufficiently strong to overcome the repulsive electric dipoles caused by charge transfer between adsorbed molecules and the silver substrate \cite{Witte2005, Torrente2007, Muller2012, Otero2017}. Second, from the molecular models depicted in the middle row of Figure~\ref{fig:SAMmodels}, the minimum spacing between oligomer edge \ce{C}-atoms is greater than \SI{0.42}{nm} in all cases, so the attractive interaction is likely to be dominated by dispersion forces \cite{Park2019}. Finally, we find that the molecular lattices have preferential orientations with respect to the substrate and appear to be commensurate with the Ag(111) lattice (see discussion of Figures~S12 and S13 in SI).

Because hydrogen is equally likely to bind to any radical site on the DTPA molecule, the synthesis of different oligomers is statistical in nature. A different flux of atomic hydrogen during DTPA deposition results in a different distribution of oligomers on the surface. We have investigated the statistics of oligomer SAMs and COF from three depositions made with identical parameters, except for the cracking filament temperature, which is set by its emission current (Figures~\ref{fig:igOnOff}D-\ref{fig:igOnOff}F). As shown in Table~\ref{table:StatisticsOfDTPA}, we find that the fraction of self-assembled monomers increases as the filament temperature (i.e., cracking rate or \ce{\Lewis{0.,H}} flux) increases, while the percentage of molecules in hexamer self-assembly decreases. The effectiveness of the reducing environment can be quantified as the number of newly-bound \ce{H} atoms per \ce{DTPA}, deduced as three minus the number of aryl--aryl bonds per monomer, and shown in Table~\ref{table:StatisticsOfDTPA} (see also Supporting Information, Figures~S4-S7).

\begin{table*}
\centering
\begin{tabular}{llclllc}
Emission & ~~Area & No. molec. & 1-mer & 2-mer & 6-mer & H/DTPA \\
\hline\hline
\SI{0.04}{\milli\ampere} & \SI{153500}{\nano\meter^2}&  \SI{31500} & 13\% & 15\% & 20\% & 1.7\\
\SI{0.10}{} & \SI{126300}{} & \SI{32200} & 36 & \phantom{0}4 & 16 & 2.0\\
\SI{1.00}{} & \SI{226400} & \SI{45500} & 67 & \phantom{0}8 & \phantom{0}- & 2.5\\
\end{tabular}
\caption{Distribution of oligomer species on the Ag(111) substrate for different cracking-filament emission currents. Fractions of molecules in monomer (1-mer), dimer (2-mer), and hexamer (6-mer) SAMs are shown, as well as the total area imaged and the number of molecules. The average number of hydrogen terminations per DTPA molecule (H/DTPA) is calculated considering all molecules, including those in mixed regions of oligomers and partial-COF that do not form well-ordered self-assembly.}
\label{table:StatisticsOfDTPA}
\end{table*}

Table~\ref{table:StatisticsOfDTPA} shows that the oligomer concentrations on the surface can be dramatically changed by the introduction of atomic hydrogen during deposition. Thus, the ability to create this reducing environment using nearly universal tools on vacuum systems, could be of interest for future studies. However, the statistical data and images presented (here and in SI) also show that the formation kinetics of the oligomers is complex. For instance, from imaging we find that hexamer SAMs tend to occur within regions of monomer SAMs and that desorption of monomers becomes highly probable at temperatures required for aryl-aryl bonding. Table~\ref{table:StatisticsOfDTPA} also exposes some of this complexity: the fraction of dimers decreases and then increases as the environment becomes more reducing (higher emission). A deeper understanding of the kinetics would enable more detailed control of the on-surface chemistry.

\subsection{Electronic Structure Evolution}
\label{sec:Electronic}

Direct comparison between on-surface experiments and the theoretical electronic structure of free-standing COFs can be challenging due to a potentially strong interaction between the metal substrate and the COF adlayer. As implied by Figure~\ref{fig:SAMmodels} (read from D to A), we can view the COF as the endpoint of a progression of molecular structures. We propose that the on-surface synthesis of these increasingly complex oligomers could lead to a more complete experimental understanding of the on-surface COF electronic structure.

\begin{figure}
  \includegraphics[width = \columnwidth]{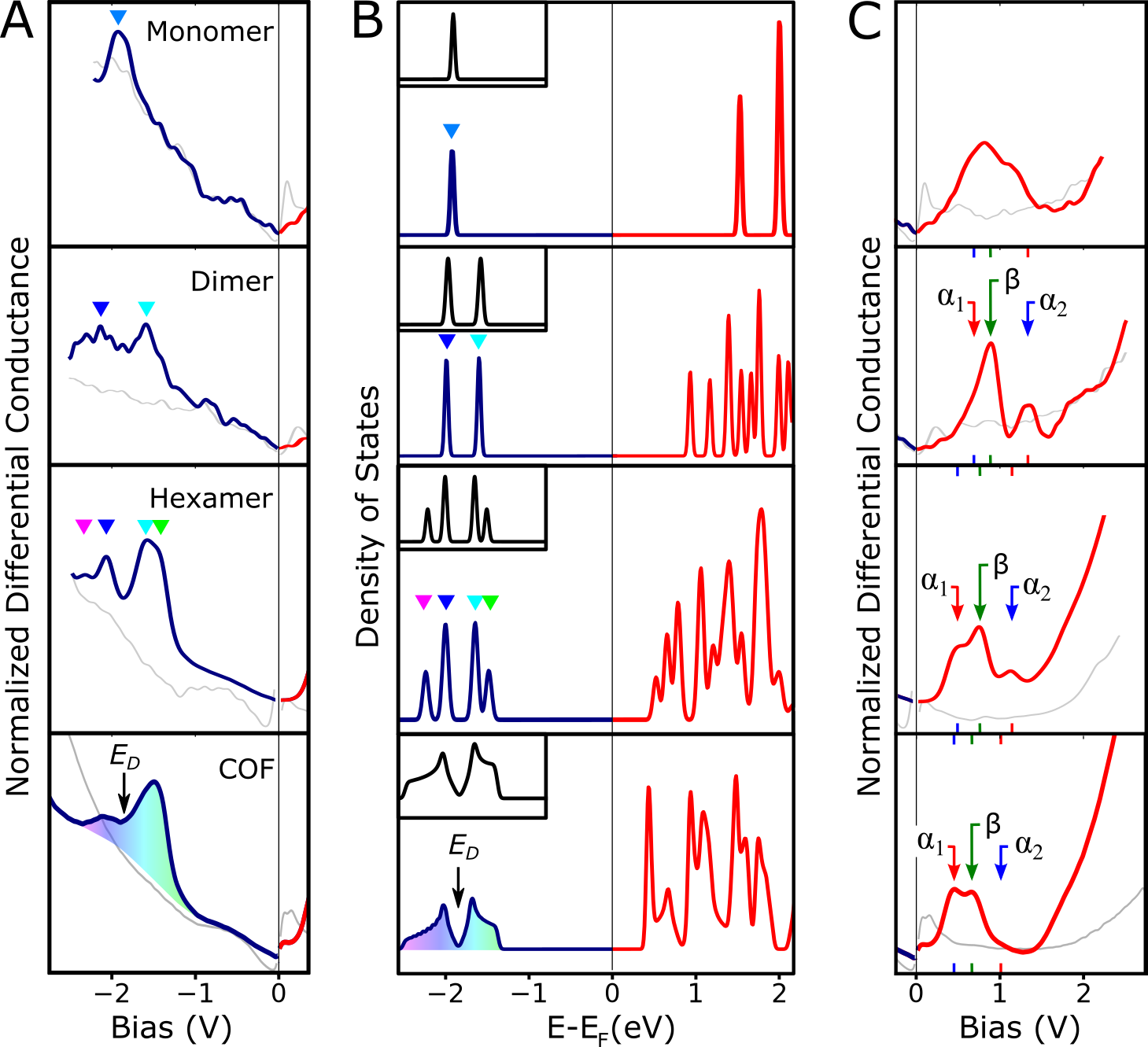}
  \caption{Comparison of the progression of occupied levels for the monomer SAM, dimer SAM, hexamer SAM, and DTPA COF. (A) Normalized differential conductance spectra (\((dI/dV)/(I/V)\)) taken via STS over monomer SAM, dimer SAM, hexamer macrocycle SAM, and DTPA COF (top to bottom) on Ag(111). Blue lines are spectra over the molecular structure and gray lines are reference spectra taken with the same tip over bare Ag(111) at the same tunnel gap impedance. The monomer and dimer spectra are individual measurements representative of a larger data set and the hexamer and COF spectra are an averaged spectrum obtained by a spectral map. (B) Freestanding DFT calculations of the occupied DoS (blue) and unoccupied DoS (red) for the monomer, dimer, and hexamer SAMs and the DTPA COF (top to bottom). The Fermi levels were set by aligning the calculated HOMOs with experimental spectra in (A). Inset boxes are tight binding calculations plotted on the same energy-axis considering a dimer and a hexamer model respectively, with the tight binding parameters $t_{nn}=\SI{-0.2}{eV}$ and $t_{nnn}=\SI{-0.012}{eV}$, where $t_{nn}$ and $t_{nnn}$ are the nearest-neighbor and next-nearest-neighbor hopping elements. (C) Continuation of the normalized differential conductance spectra from (A) but over the unoccupied electronic states. Labels $\alpha_1$, $\alpha_2$, and $\beta$ mark bias voltages of the peak positions in Figure~\ref{fig:Unfilled}. The SAM data was acquired at \SI{77}{\kelvin}, while the COF data was acquired at \SI{4.7}{\kelvin}.}
  \label{fig:dIdVstack}
\end{figure}

Figure~\ref{fig:dIdVstack}A (Figure~\ref{fig:dIdVstack}C) displays experimental filled-state (unfilled-state) STS obtained within SAMs of monomers, dimers, hexamer macrocycles, and a COF island. In Figure~\ref{fig:dIdVstack}B we show the density-of-states (DoS) calculated at the DFT-PBE level for the freestanding oligomers and COF. For the highest occupied molecular orbital (HOMO) levels, both experiment and theory show a simple trend of peak splittings, which is consistent with the intuitive notion of a linear combination of molecular orbitals (LCMO)---one MO contributed by each monomer \cite{Jiang2021}. A simplified effective model of interacting MOs can be implemented using a tight-binding model with a single orbital per site \cite{Thomas19a} (see Figure~S14 and accompanying text). The insets to Figure~\ref{fig:dIdVstack}B show that this model reproduces the basic STS and DFT results, including the symmetry-expected 1:2:2:1 degeneracies of the hexamer DoS peaks and the slight asymmetry in the peak energies for the hexamer and COF bands (within the model, the asymmetry is due to next-nearest neighbor interactions). 

Over the same energy range as the hexamer HOMO levels, the calculated COF spectrum (bottom panel of Figure~\ref{fig:dIdVstack}B) shows similar spectral characteristics, except that individual levels broaden into bands. Experimentally, the correspondence is less obvious due to the energy-dependent background in STS (light-gray lines show STS over the silver surface). The amplitude and width of STS features may also be affected by the finite size of the COF island, lattice defects, and energy-dependent mechanisms such as the hole lifetime. However, tracking the progression of electronic levels to their COF endpoint enables an unambiguous experimental identification of the HOMO bands and particularly the Dirac point ($E_D$) for these bands, which clearly lies at the minimum STS intensity near \SI{-1.9}{eV} (see also the band structures calculated at the DFT-PBE level in Figure~S15). In Figure~\ref{fig:dIdVstack}B, the DFT spectra have been shifted to align with the experimental spectra by matching $E_D$ and other prominent features.
\begin{figure}
  \includegraphics[width=\columnwidth]{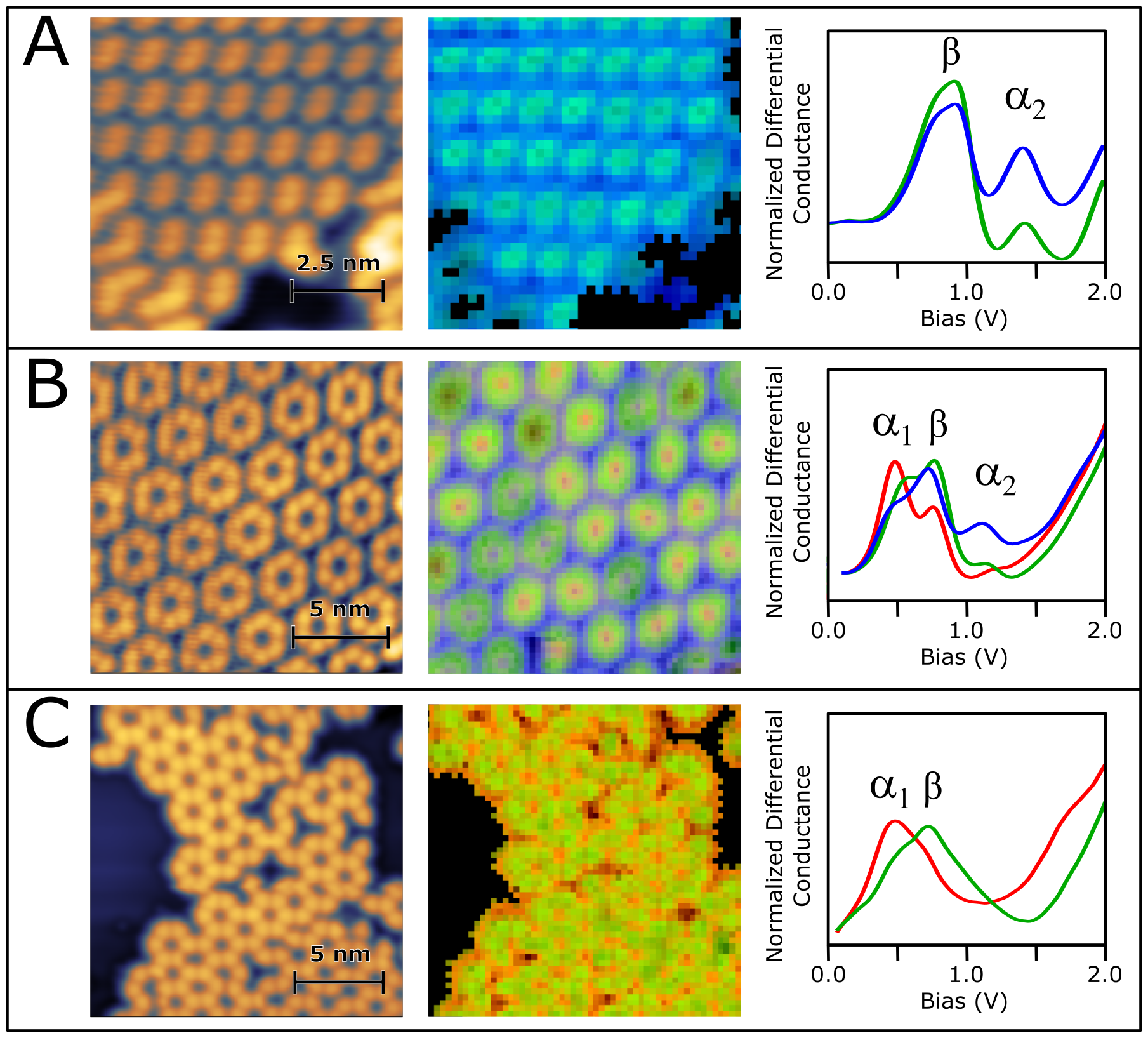}
  \caption{Each row shows the results of spectral maps taken over a
    dimer SAM (A), hexamer SAM (B), and DTPA COF (C). In each row the
    first image is the STM topograph. The RGB values in the second
    image are the normalized Euclidean distance from the spectra taken
    at each pixel location to the corresponding spectra in the third
    column. Here the Euclidean distance refers to the vector 2-norm
    computed in the vector space spanned by the bias voltages. These
    values are used to visualize the variation of the spectrum taken
    at each point in the map to the averaged red, green, or blue
    spectrum in the third column. The plots in the third column are
    the normalized differential conductance spectra of the unfilled
    states used as the origin to compute the Euclidean distance and
    were obtained by averaging the spectra over the pores (red), the
    molecules (green), and the interstitial regions (blue). The
    spectra were grouped together by K-means clustering (see
    Figure~S20). The dimer SAM and COF spectral maps were acquired at
    \SI{4.7}{\kelvin}, while the hexamer data was acquired at
    \SI{77}{\kelvin}.}
  \label{fig:Unfilled}
\end{figure}

In contrast to the filled states, the evolution of the lowest unoccupied molecular orbitals (LUMOs) is complicated from the outset by the presence of closely spaced MOs in the monomer, as shown in the DFT results of Figure~\ref{fig:dIdVstack}B. Effective tight-binding models using a basis set of $\pi_{x}, \pi_{y},$ and $\sigma$ orbitals have previously illuminated interesting aspects of 2D COF bandstructure \cite{Ni2022,Jiang2021,Wu2008}, but such an approach is less intuitive for this system with several closely spaced MOs. More importantly, the experimental spectra from unfilled electronic states (Figure~\ref{fig:dIdVstack}C) bear little resemblance even to our DFT results. We take this as an indication that---over the LUMO energy range---the influence of the substrate is significant, which is undoubtedly due to the presence of the Ag(111) surface state above \SI{-55}{\milli\eV}. Deciphering the observed spectral peaks in this situation is difficult, but it is facilitated by having regular arrays of successively larger oligomers, where the influence of the surface state can be isolated to some extent.

Spectra taken over the SAMs and COF consistently show distinct local maxima within the range \SIrange{0}{1.5}{eV}. The peak energies, widths, and amplitudes vary from monomer through hexamer, but become similar for the hexamer and COF. For discussion, $\alpha_1$, $\beta$, $\alpha_2$, and accompanying lines/tics in Figure~\ref{fig:dIdVstack}C mark peaks of potentially similar origin, as determined by their spatial distribution. 

The left column of Figure~\ref{fig:Unfilled} displays topographs taken over regions of dimer SAM (A), hexamer SAM (B), and COF (C), with data from corresponding spectral maps shown in the middle and right columns. The K-means clustering algorithm \cite{MacQueen1967} was used to group 64x64 spectra from each map into four (dimers and COF) or five (hexamers) clusters, based on their spectral features within \SIrange{0}{2}{\volt} (see Figure~S16 for additional details). Of these clusters, one always corresponds to spectra from the Ag(111) substrate and another to spectra displaying transient ``tip-switches'' or other anomalies; these two clusters are not considered in further analysis. Spectra within each of the remaining clusters (labeled and colored as red, green, or blue) are averaged and shown in the right column of Figure~\ref{fig:Unfilled} (these average spectra comprise the ``origin vector'' to the center of each cluster, where the size of the vector space is the number of bias-voltage samples). Comparing with Figure~\ref{fig:dIdVstack}C, it is clear that the K-means unsupervised machine learning picks out spectra that are dominated by different peaks among $\alpha_1$ (red), $\beta$ (green), and $\alpha_2$ (blue).  We visualize the distribution of each cluster (roughly speaking, each spectral peak) over the imaged regions using color intensity to  represent the normalized Euclidean distance (vector 2-norm) of individual spectra from their respective cluster centers. The resulting images are shown in the middle column of Figure~\ref{fig:Unfilled}, with different cluster data represented by intensities on different color channels of the same image (colors correspond to spectra in the right column). 

The red-cluster spectra from the hexamer SAM and COF have $\alpha_1$ as the tallest peak and are located predominantly over the pores. This is inconsistent with the calculated charge density maps for the freestanding COF LUMO (see Figure~S17) or indeed, any COF level. Peak $\beta$ is present in all cluster spectra but has the highest relative intensity in the green cluster, which overlies the molecular backbone for all three structures (dimer SAM, hexamer SAM, and COF). Finally, the $\alpha_2$ peak attains its highest intensity in the blue-cluster spectra which are found in the interstitial regions of the SAMs. The $\alpha_2$ peak essentially vanishes for the COF, which has no interstitial region, and is also absent for an isolated hexamer spectrum, a further indication that it relates to the SAM formations. 

Comparing the spatial distribution of spectral data from the COF and different SAMs facilitates the separation of features dominated by molecular states versus those dominated by the substrate surface state. With their prevalence over regions of bare substrate, we associate the red and blue K-means clusters with metal-dominated features ($\alpha_1$ and $\alpha_2$ peaks), whereas the green cluster corresponds to molecule-dominant LUMO states. ``Quantum corral'' confinement of noble-metal surface states has been known for many years \cite{Crommie1993} and has more recently been studied and modeled for metal-organic frameworks and organic SAMs \cite{Kepcija2015, Yang2018, Zhou2020, Piquero-Zulaica2017}. A straightforward calculation for a circular quantum well of radius \SI{0.94}{\nano\metre} (the COF pore diameter) results in a confinement energy of \SI{0.54}{\eV} (see Supporting Information discussion ``Electron confined to a circle in two dimensions''). This is consistent with the energy of peak $\alpha_1$ for both the hexamer and the COF, so we have some confidence that its source is resonant scattering of the Ag(111) surface state within the pores. Similarly, the spectral maps of Figure~\ref{fig:Unfilled} show that peak $\alpha_2$ is largely a consequence of a surface state resonance in the interstitial regions of the dimer and hexamer SAMs, for which the intermolecular spacings are approximately the same. This conclusion is supported by a plane-wave model of surface-state electrons propagating in a 2D lattice of model Gaussian potentials, each situated under an atom of the polymer (see Figure~S18 and accompanying text). The model predicts a resonance at around \SI{1.0}{\eV}, which is close to the observed energy of $\alpha_2$. Finally, we note that the absence of peak $\alpha_2$ for the COF spectrum of Figure~\ref{fig:dIdVstack} is expected in this interpretation since no interstitial region exists. We interpret a weak shoulder near \SI{1}{eV} as a second-order pore resonance, which is also reproduced in these scattering models. 

Peak $\beta$ may also be influenced by the metal surface state, but its spatial distribution implies that it derives mainly from the adsorbed molecular layer. Identifying $\beta$ as the LUMO peak, we can extract a more accurate bandgap than otherwise possible. For our COF experiments, the derived value of the bandgap, \SI{1.85\pm 0.1}{eV}, is nearly identical to DFT results for the freestanding COF (Figure~\ref{fig:dIdVstack}B). (This close agreement may be fortuitous; we expect that the calculated bandgap would decrease somewhat with the inclusion of a Ag(111) substrate beneath the COF.) 

\section{Conclusions}
\label{sec:Conclusion}

This work establishes the disruptive effect of atomic hydrogen during the on-surface synthesis of covalent organic frameworks, with quantitative confirmation of the hydrogenated products that impede Ullmann coupling. We find that even a remote cracking filament and modest hydrogen pressure can create a sufficiently reducing environment to dramatically inhibit \ce{C-C} bonding in ambient high vacuum, and postulate that in ultrahigh vacuum the same effect could limit the size and quality of 2D COF crystals. Therefore, to achieve large-area 2D COFs in vacuum, the level of background hydrogen must be minimized, and bare filaments should be cold during COF synthesis. These findings will affect the design of heating stages, chamber construction, and pump selection for dedicated 2D COF facilities. Our research also shows that these inhibitory processes can be controlled and exploited to produce---from a single monomer precursor---a progression of oligomers (monomers, dimers, hexamers, and to a lesser extent, 3-,4-,5-mers) organized into self-assembled monolayers, which have their own utility and scientific interest. Other 2D organic networks, with analogous oligomer sequences, could benefit from the deposition and analysis techniques demonstrated here. Finally, we expect that deposition in this controlled reducing environment can be used to manage the length distribution of surface-synthesized one-dimensional polymers, ensuring a known hydrogen termination and allowing the exploration of length-dependent properties such as molecular conductance \cite{Lafferentz2009} and electronic structure \cite{Mishra2021}. 

\section{Methods}
\label{sec:Methods}

\paragraph{Precursor synthesis and depositions.} \ce{(Br_3)DTPA} and \ce{(Br_3)OTPA} precursor molecules were synthesized as per the reported procedures \cite{Kuratsu2005, Suzuki2012, Fang2012}. DTPA depositions were conducted in the baked-out load-lock with a pressure of \SIrange[range-units=single,range-phrase=--]{2}{3e-8}{\milli\bar}. DTPA precursor molecules are evaporated from a home-built Knudsen dual-crucible evaporator, made of boron nitride, onto a UHV sputter-cleaned Ag(111) substrate approximately \SI{0.4}{\metre} away. The Ag(111) thin films, grown on cleaved Mica \cite{Baski1994}, were prepared by cycles of sputtering and annealing in ultrahigh vacuum (UHV) prior to each deposition. The substrate is held at a constant temperature during deposition and for 5--10 minutes after the deposition is complete. The sample is then moved into the UHV chamber (\(<\SI{1e-10}{mbar}\)) and onto the cooled STM stage. Depositions in the \ce{\Lewis{0.,H}}-rich environment have 99.9999\% purity \ce{H_2} continuously streamed into the load-lock chamber via an attached variable leak valve, maintaining a specified pressure. Depositions with the ``cracking source on'' were performed with a Bayard-Alpert ion gauge equipped with a thoriated iridium filament operating with a hot filament emission current of \SI{0.04}{mA}, \SI{0.1}{\milli\ampere} or \SI{1}{\milli\ampere} measured at the grid generated by passing \SI{4.0}{\ampere}, \SI{4.1}{\ampere} and \SI{4.4}{\ampere} respectively through the filament. The temperature of the filament is maintained by the ion gauge controller using the emission current as feedback.

Deposition parameters vary for Figure~\ref{fig:SAMmodels}. Figure~\ref{fig:SAMmodels}A had the ion gauge filament off throughout the duration the sample was in the load-lock. The background pressure was less than \SI{2.3e-8}{\milli\bar} for the 12 minute deposition at a deposition rate of 0.05 ML/min (molecular monolayers per minute) onto a \SI{300}{\celsius} Ag(111) substrate. For Figure~\ref{fig:SAMmodels}B, the sample was held in the load-lock chamber for 15 minutes with the ion gauge on and a background pressure of \SI{2.4e-8}{\milli\bar} prior to the deposition. The deposition was then conducted with the ion gauge off for 11 minutes at a deposition rate of 0.05 ML/min onto a \SI{300}{\celsius} Ag(111) substrate. Figure~\ref{fig:SAMmodels}C exposed the sample to the ion gauge for 20 minutes prior to deposition in the load-lock with a background pressure of \SI{4.7e-8}{\milli\bar}. The ion gauge was left on during the 8.25-minute deposition at a rate of 0.065 ML/min onto a \SI{315}{\celsius} Ag(111) substrate. Finally, Figure~\ref{fig:SAMmodels}D had the sample exposed to the ion gauge for 30 minutes prior to deposition in the load-lock with a background pressure of \SI{2.8e-8}{\milli\bar}. The ion gauge remained on during the 10.75-minute deposition at a rate of 0.045 ML/min onto a \SI{300}{\celsius} Ag(111) substrate.

The densities of the monomer and dimer assemblies are $7.4\times 10^{13}$ molecules per \si{cm^2}, while the densities of hexamer assembly and COF are $8.6\times 10^{13}$ and $7.6 \times 10^{13}$ molecules per \si{cm^2}, respectively.

\paragraph{STM.}
Measurements were performed with a CreaTec LT-STM operating in UHV (\(< \SI{1e-10}{mbar}\)) at \SI{77}{\kelvin}. All STM images were taken in constant current mode with \ce{W} tips etched in the lab and cleaned via \latin{in situ} field emission over a clean Ag(111) substrate and subsequent conditioning, typically via nanoscale contact with the Ag substrate. All STM conductance spectra presented were acquired as part of \(64\times 64\) spectroscopic maps, where $dI/dV$ spectra are acquired after opening the feedback loop at each grid-point. The STM operates in constant current mode between spectra. A typical acquisition time for a topograph is 30 seconds and the specific tunneling conditions for each image is listed below. We used Gwyddion \cite{Necas2012} to process images (e.g., line leveling, color range, Gaussian filter). Tunneling spectra were analyzed by a home-built Python tool. \(dI/dV\) spectra are normalized by taking \((dI/dV)/(I/V)\), where \(I\) is generated by numerically integrating a heavily smoothed \((dI/dV)\) signal, and data points very close to the Fermi level (zero sample bias) are removed.  \(dI/dV\) data is obtained using the internal digital lock-in amplifier integrated into the CreaTec hardware/software, which directly modulates the bias voltage and demodulates the current signal.  In this work the lock-in frequency is \SI{1111}{\hertz} and the lock-in amplitude is \SI{45}{mV_{pp}}.

The topographs in Figure~\ref{fig:igOnOff} were acquired at (Sample bias, tunnel current): (A) \SI{-1.95}{V}, \SI{58}{\pico\ampere}; (B) \SI{2.07}{V}, \SI{63}{\pico\ampere}; (C) \SI{1.80}{V}, \SI{87}{\pico\ampere}; (D) \SI{-1.13}{V}, \SI{69}{\pico\ampere}; (E) \SI{-0.85}{V}, \SI{120}{\pico\ampere}; (F) \SI{1.35}{V}, \SI{16}{\pico\ampere}.

The topographs in Figure~\ref{fig:SAMmodels} were acquired at: (A) \SI{-0.28}{V}, \SI{150}{\pico\ampere}; (B) \SI{-0.43}{V}, \SI{96}{\pico\ampere}; (C) \SI{-1.83}{V}, \SI{63}{\pico\ampere}; (D) \SI{-2.03}{V}, \SI{120}{\pico\ampere}.

The spectral map data presented in Figure~\ref{fig:Unfilled} were acquired in an open loop mode with the tunneling setpoints: (A) \SI{0.56}{V}, \SI{220}{\pico\ampere}; (B) \SI{-0.50}{V}, \SI{120}{\pico\ampere}; (C) \SI{-0.52}{V}, \SI{26}{\pico\ampere}.

\paragraph{TOF-SIMS.}
Samples were prepared by \latin{in-situ} deposition of monomers onto a Au(111) thin film substrate to a uniform coverage of \SI{~1}{ML} as determined by STM. A characterized sample was then transferred to a TOF-SIMS instrument (IONTOF 5-300), requiring a \SI{< 10}{min.} exposure to atmosphere. After transfer and subsequent vacuum pump-down, the samples were analyzed using a primary ion beam of \ce{Bi_3^{2+}} cluster ions with an energy of \SI{50}{keV} and a secondary ion extraction voltage of \SI{2}{kV}. The cumulative spectra shown in Figure~\ref{fig:TOFSIMS} were acquired during raster scans of the primary ion beam over $\SI{500}{\mu m} \times \SI{500}{\mu m}$ regions of the samples.

\paragraph{DFT.}
Density functional theory (DFT) calculations were performed with the projector augmented wave (PAW) method \cite{Blochl1994}, as implemented in the Vienna Ab initio Simulation Package (VASP) \cite{Kresse1996, Kresse1996a}. The generalized gradient approximation (GGA)/Perdew-Burke-Ernzerhof (PBE) functional\cite{Perdew1996} including the Grimme dispersion correction (DFT-D3)\cite{Grimme2010} were used for geometry optimizations and electronic structure calculations. In the geometry optimization of the \ce{2CH_3}-bridged triphenylamine COF monolayer, the cut-off energy was set to \SI{500}{\electronvolt} and the $\Gamma$-centered $9\times9\times1$ k-grids with Monkhorst-Pack scheme were adopted. Both the lattice parameters and atomic coordinates were fully relaxed until the force on each atom and the variation in total energy were smaller than \SI{0.01}{\electronvolt \per \angstrom} and \SI{1e-6}{\electronvolt}, respectively. For the single-point electronic density of states (DOS) calculations, a $21\times21\times1$ k-mesh was used. In the geometry optimizations of the DTPA monomer, dimer, and hexamer, the same convergence criteria as the COF monolayer was adopted using only the $\Gamma$ point. For the interface system of the \ce{CH_2}-bridged triphenylamine COF on a five-layer Ag slab, the bottom three Ag layers were fixed while the top two Ag layers and the COF were fully relaxed. K-meshes of $3\times3\times1$ and $9\times9\times1$ were used for the geometry optimizations and electronic structure calculations, respectively.

\begin{acknowledgement}

  The authors thank W. Henderson for training and assistance with TOF-SIMS measurements and X. Ni, W. R. Dichtel, M. F. Crommie, G. Wang, and A. D. Vira for helpful discussions and advice. This work was funded by the United States Army Research Office under grant number W911NF-15-1-0447 (MURI). Partial support from NASA-SSERVI (Cooperative Agreement No. NNA17BF68A) and seed funding from Georgia Tech STAMI are also acknowledged. Experiments were performed in part within the Materials Characterization Facility (MCF) at Georgia Tech.  The MCF is jointly supported by the GT Institute for Materials (IMat) and the Institute for Electronics and Nanotechnology (IEN), which is a member of the National Nanotechnology Coordinated Infrastructure supported by the National Science Foundation (Grant ECCS-2025462).

\end{acknowledgement}

\begin{suppinfo}
Experimental procedures and supporting information are available online.

Geometry of load-lock deposition chamber, STM images of DTPA deposition at different substrate temperatures, experimental data demonstrating that the hot filament in the ion gauge is the primary cracking source, experimental data showing the varying surface composition of DTPA species with varying cracking filament temperature, STM images of DTPA deposition on Au(111) substrate, additional ToF-SIMS data with Br-terminated and H-terminated DTPA, STM images of OTPA deposition on Au(111) substrate, ToF-SIMS data obtained from deuterium-substituted OTPA, description of the commensurate molecular DTPA overlayers on Ag(111) surface, DFT calculations of the adsorption configurations on Ag(111) surface, description of the tight binding calculations, bandstructures of freestanding DTPA COF obtained by DFT, additional description of the STS spectral map analysis in Figure \ref{fig:Unfilled}, partial charge density maps of the freestanding DTPA COF obtained by DFT, description of electron confined in a pore, and description of the electron plane wave expansion modeling.  
\end{suppinfo}

\bibliography{references}

%%% START arxiv: TOC Graphic squeezed after bibliography
\vspace{2em}
%\phantom{99}\includegraphics[height = 3.5 cm]{"figures/Graphical TOC V2D.png"}
\phantom{99}\includegraphics[height = 3.5 cm]{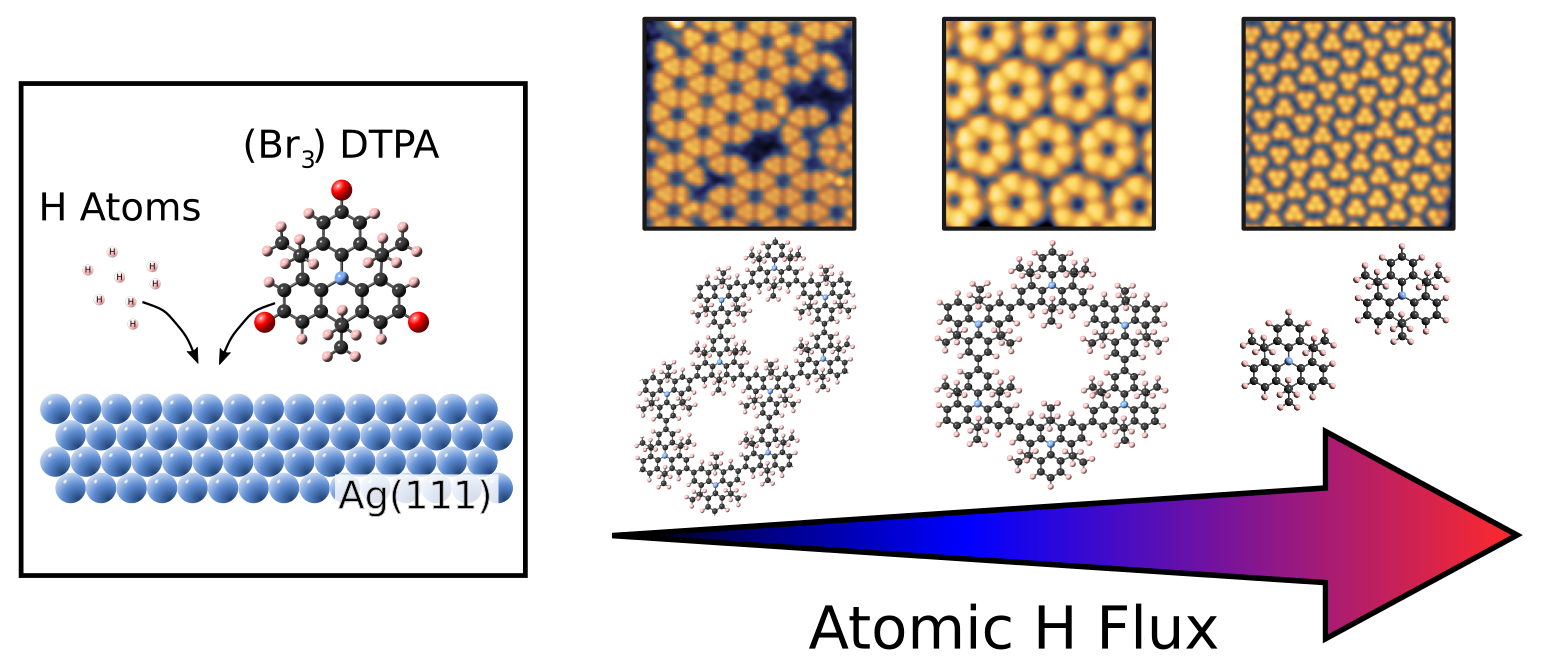}
%%% END arxiv

\end{document}

%%%%%%%%%%%%%%%%%%%%%%%%%
\section{Load-lock Deposition Geometry}
\label{sec:Depo}

\begin{figure}[H]
  \includegraphics[width=.85\columnwidth]{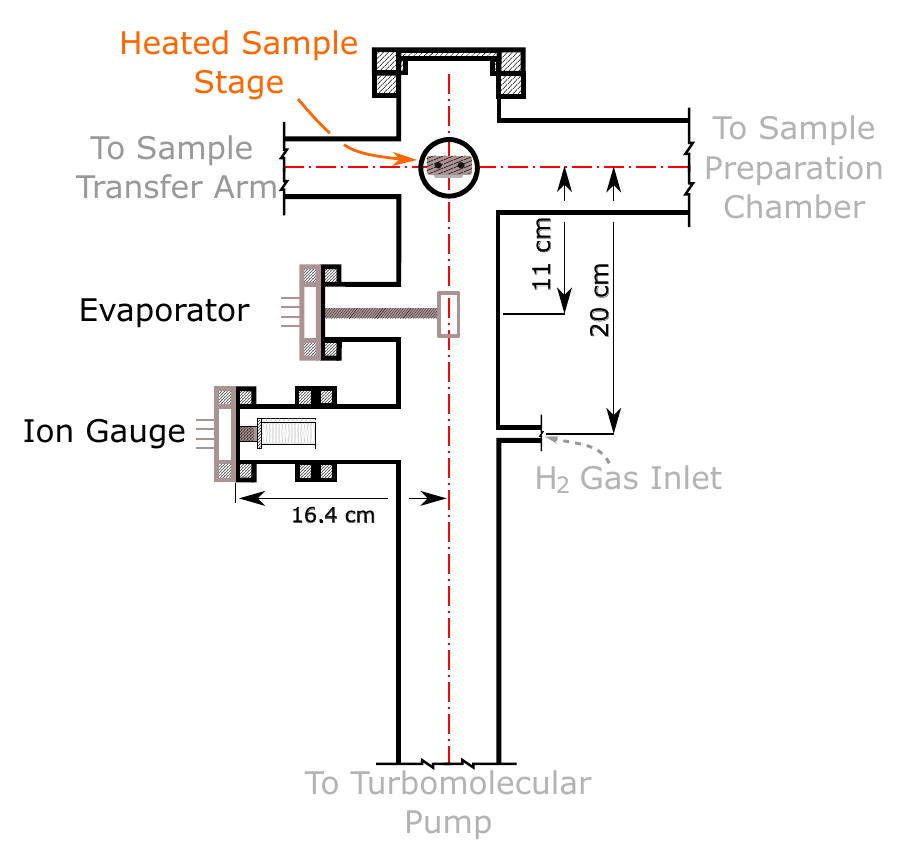}
  \caption{Schematic of the turbo-pumped sample-introduction chamber (load-lock) used for molecular depositions. Molecules originate from the dual crucible evaporator positioned \SI{11}{\centi\metre} from the heated sample stage. The \ce{H2} cracking filament is the primary filament on the ion gauge positioned below the sample stage and evaporator.}
\label{fig:Depo}
\end{figure}

%%%%%%%%%%%%%%%%%%%%%%%%%%%%%%
%\newpage
\section{Substrate Temperature and Monomer Desorption}
\label{sec:Desorp}

\begin{figure}[H]
  \includegraphics[width=.75\columnwidth]{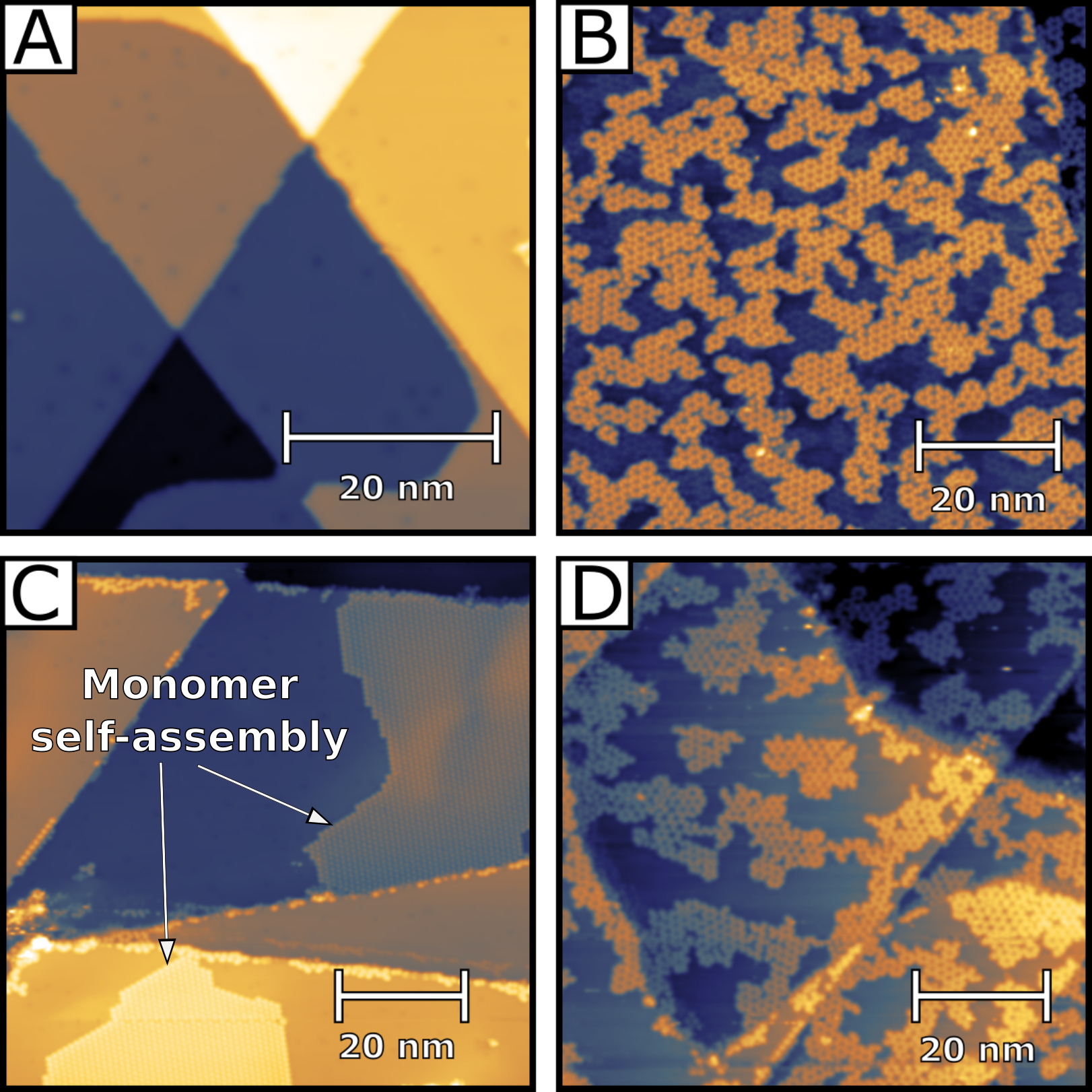}
  \caption{Topographs of DTPA depositions under different conditions on Ag(111). (A) \ce{H2} backfilled to \SI{2.2E-6}{\milli\bar} with cracking filament on (0.1 mA emission current) and \SI{543}{\kelvin} substrate temp. (B) HV deposition with cracking filament off and \SI{543}{\kelvin} substrate temp. (C) \ce{H2} backfilled to \SI{2.2E-6}{\milli\bar} with cracking filament on (0.1 mA emission current) and \SI{523}{\kelvin} substrate temp. (D) HV deposition with cracking filament off and \SI{523}{\kelvin} substrate temp. Imaging conditions: (A) \SI{1.82}{\volt}, \SI{24}{\pico\ampere}, (B) \SI{-2.07}{\volt}, \SI{29}{\pico\ampere}, (C) \SI{2.07}{\volt}, \SI{63}{\pico\ampere} and (D) \SI{-1.80}{\volt}, \SI{140}{\pico\ampere}.}
\label{fig:Desorb}
\end{figure}

Desorption of the hydrogen terminated DTPA monomers was found to occur at lower substrate temperatures than for other oligomers (Figure~\ref{fig:Desorb}). The desorption was observed by a decrease in molecular coverage after depositing in a \ce{H_2} backfilled environment with the cracking filament on (which leads to approximate total hydrogenation of the debrominated DTPA sites). At \SI{543}{\kelvin} substrate temperature no molecules are present on the surface (Figure~\ref{fig:Desorb}A), whereas the control deposition (filament off) has $\approx70\%$ monolayer coverage (B). At \SI{523}{\kelvin} substrate temperature the filament-on deposition has $\approx32\%$ monolayer coverage (C) whereas the control deposition had $\approx65\%$ monolayer coverage (D). 

\section{Verification of the hot filament as the primary source of atomic hydrogen}
\label{sec:IGFil}

\begin{figure}[H]
  \includegraphics[width=\columnwidth]{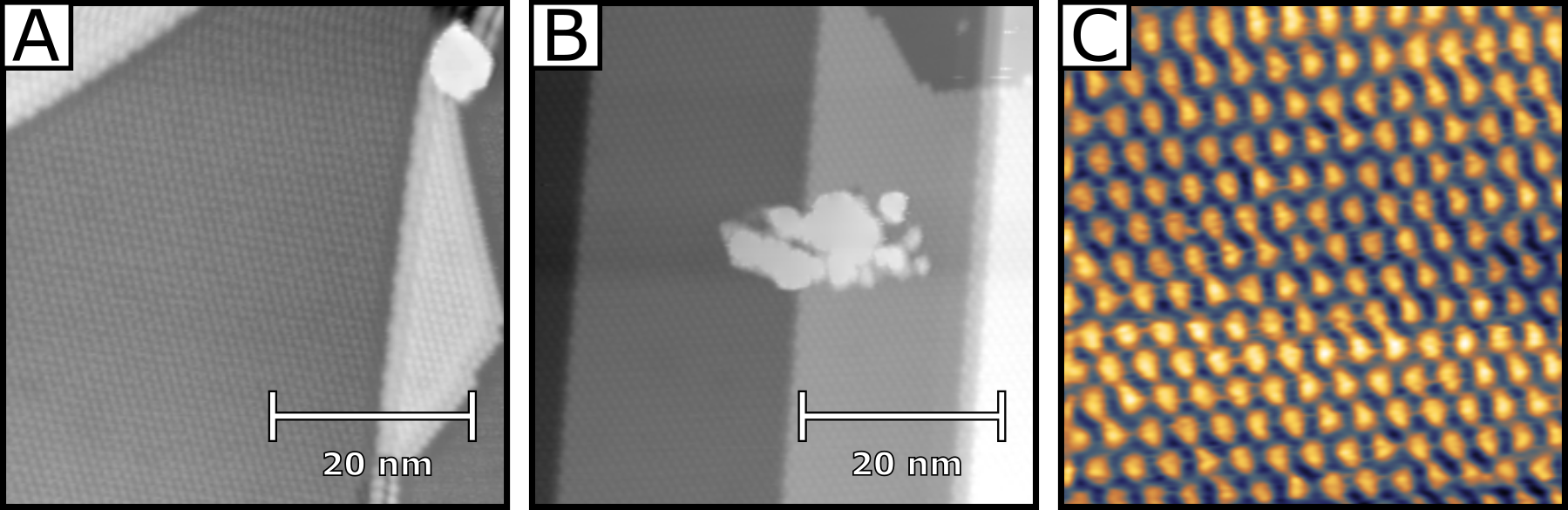}
  \caption{STM topographs of a DTPA deposition onto Ag(111) performed in a \ce{H_2} backfilled chamber (\SI{2.2E-6}{\milli\bar}) and the heated ion gauge filament. The filament was heated directly by an AC current of \SI{4.15}{\ampere} (approximate filament current for \SI{0.1}{\milli\ampere} emission current) without a grid voltage. This deposition showed only large (\SIrange{50}{100}{\nano\metre}) islands of monomer SAMs. (A,B) Examples of large monomer SAMs. (C) Zoom of monomer SAM region. Imaging parameters: (A,B) \SI{1.18}{\volt}, \SI{28}{\pico \ampere} and (C) \SI{-1.18}{\volt}, \SI{22}{\pico \ampere}.}
  \label{fig:IGFil}
\end{figure}

To verify that the most significant source of \ce{\Lewis{0.,H}} is the heated filament of the ion gauge and not the generated ions or electrons, Figure~\ref{fig:IGFil} shows a deposition of DTPA in the presence of \ce{H_2} with only a heated ion gauge filament (no grid or filament bias voltages). This resulted in a surface covered with only monomer SAM; the same result as obtained by deposition in \ce{H_2} with the ion gauge operating normally (i.e., with grid and filament bias voltages).

%%%%%%%%%%%%%%%%%%%%%%%%%%%%%%
%\newpage
\section{DTPA on Ag(111): Surface composition with varying cracking filament temperature}

The deposition of \ce{(Br_3)DTPA} onto a heated Ag(111) substrate in the presence of a source of atomic hydrogen results in a mixture of DTPA species with varying degrees of hydrogenation, due to the stochastic nature of the termination due to atomic hydrogen. When a hot filament is absent, the surface shows the formation of covalently-bonded islands even in the presence of molecular hydrogen as shown in Figure \ref{fig:Desorb}B,D.

Performing the depositions with an excess of atomic hydrogen results in virtually all the molecules being fully hydrogenated, as the surface is covered only by the monomer SAM, for example, Figure \ref{fig:Desorb}C, \ref{fig:IGFil}C. Depositions with lower amounts of hydrogen produce mixtures of oligomers. A fraction of these various oligomeric species form well-ordered self-assemblies. 

To illustrate this, we have performed three depositions with identical molecular deposition rates, background pressures (\SI{3e-8}{mbar}) and substrate temperatures (\SI{503}{\kelvin}), for the same duration of time (10 minutes) while varying the temperature of the cracking filament. The temperature of the cracking filament was varied by changing the thermionic emission current of the filament. Depositions were performed with emission currents of \SI{0.04}{\milli\ampere}, \SI{0.1}{\milli\ampere} and \SI{1}{\milli\ampere}, and the surfaces were scanned using STM to identify the differences in the sample composition.

\begin{figure}[H]
  \includegraphics[width=\columnwidth]{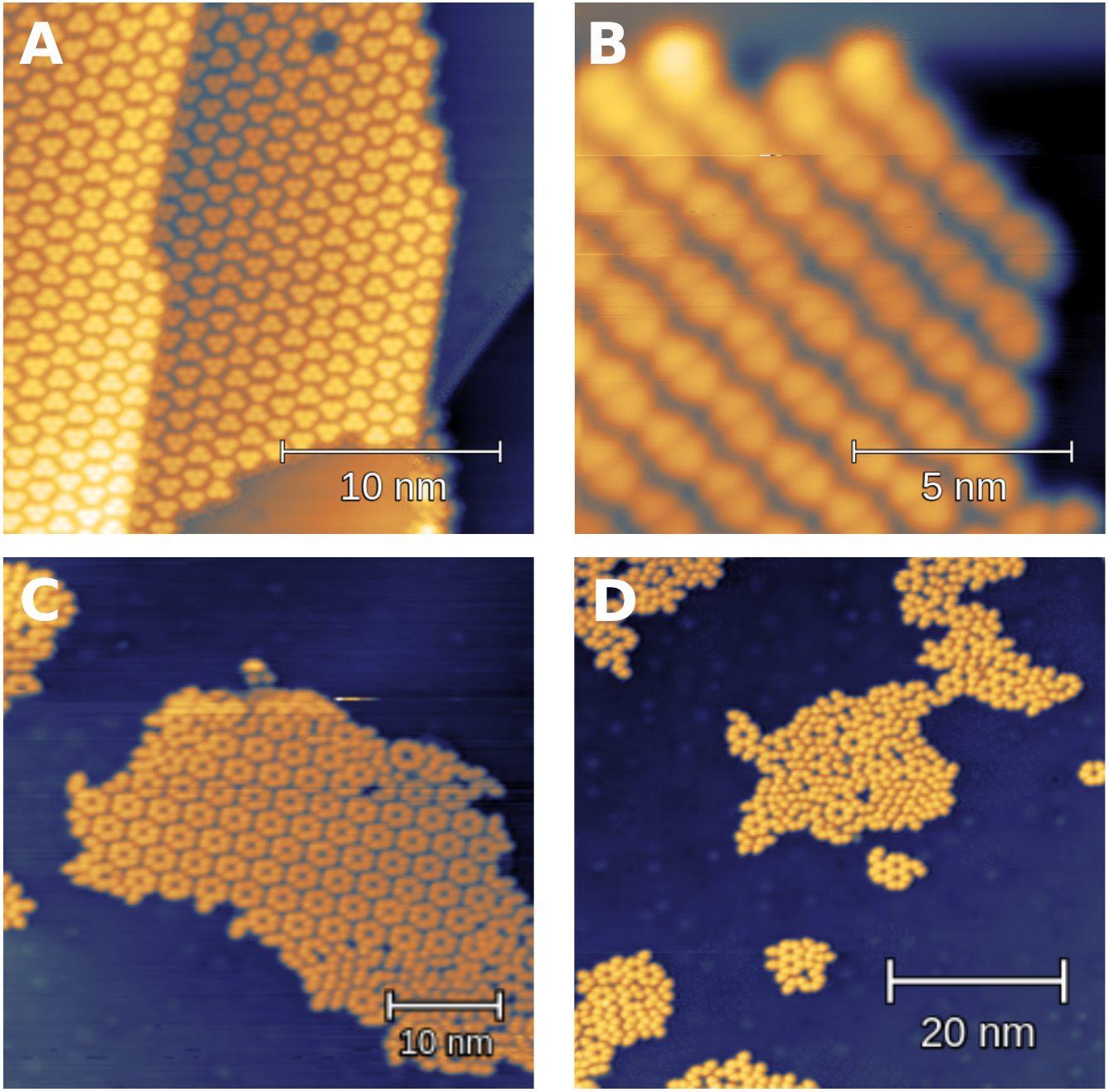}
  \caption{Species typically found on the Ag(111) surface after deposition of DTPA in the presence of atomic hydrogen. The commonly found ordered self-assemblies are that of (A) monomers, (B) dimers and (C) hexamers. In addition to the ordered self-assemblies, aggregates of oligomers without well-defined self-assembly (D) can also be found. Imaging parameters: (A) \SI{-0.97}{\volt}, \SI{130}{\pico\ampere} (B) \SI{-1.44}{\volt}, \SI{25}{\pico\ampere} (C) \SI{-0.86}{\volt}, \SI{110}{\pico\ampere} and (D) \SI{-2.00}{\volt}, \SI{150}{\pico \ampere}. }
  \label{fig:Surface_examples}
\end{figure}

The species found on the surface can be broadly categorized as self-assembled monomers (Figure~\ref{fig:Surface_examples}A), dimers (B), hexamers (C), and mixed oligomers (D). 

The results of the depositions broadly follow the expected trend of higher filament temperature causing a greater degree of hydrogenation of the DTPA species. 

\begin{figure}[H]
  \includegraphics[width=\columnwidth]{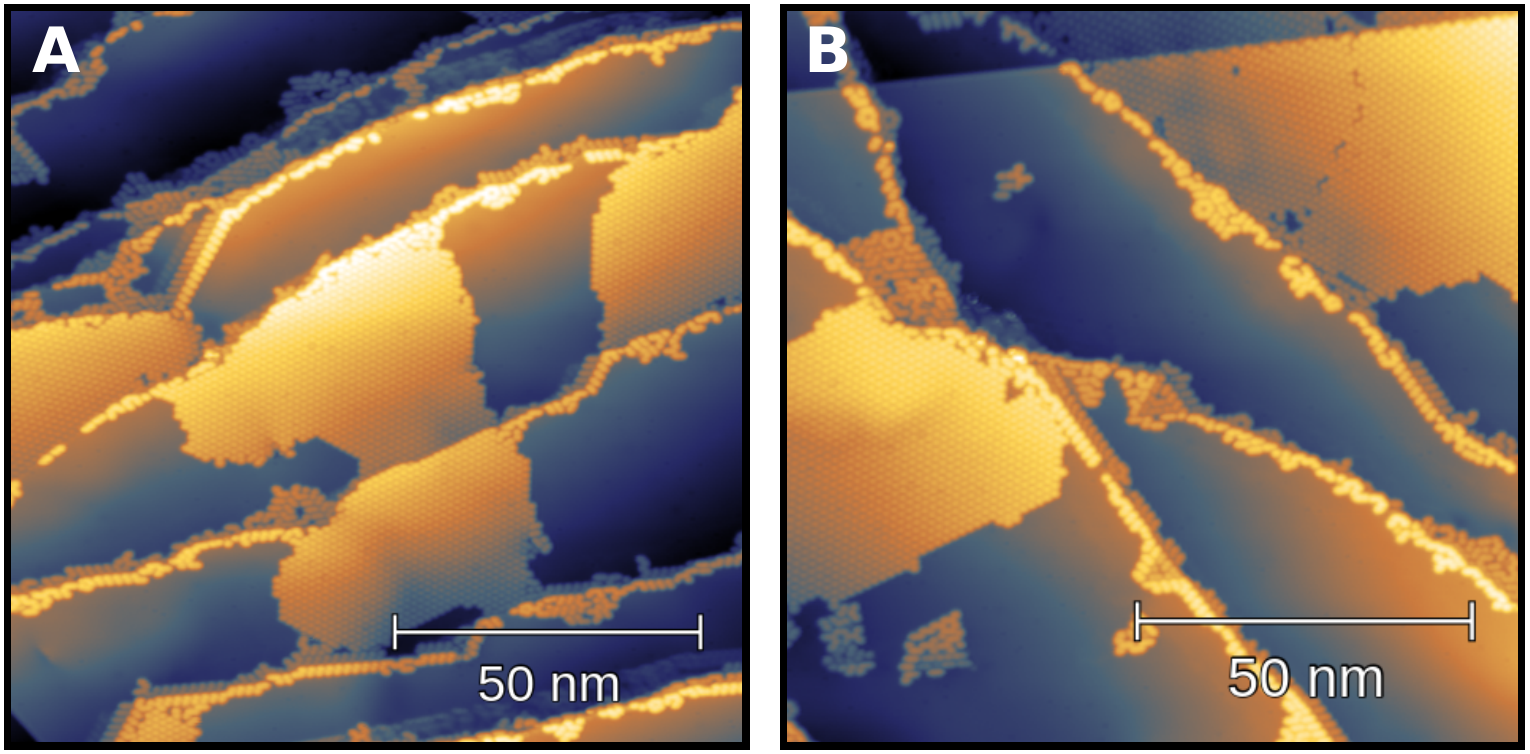}
  \caption{Deposition performed with a filament emission current of \SI{1}{\milli\ampere}. The surface is dominated by monomer self-assembly, with small islands of dimers and mixed oligomers on the surface and step-edges. Imaging parameters (for both (A) and (B)): \SI{1.35}{\volt}, \SI{16}{\pico\ampere}.}
  \label{fig:1mA}
\end{figure}

%\newpage
\textbf{1 mA emission:} This results in a surface that is dominated by monomer self-assembly, along with a small proportion of dimer self-assembly and mixed oligomers (Figure \ref{fig:1mA}). Hexamer self-assembly was not seen when scanning this sample.

\begin{figure}[H]
  \includegraphics[width=\columnwidth]{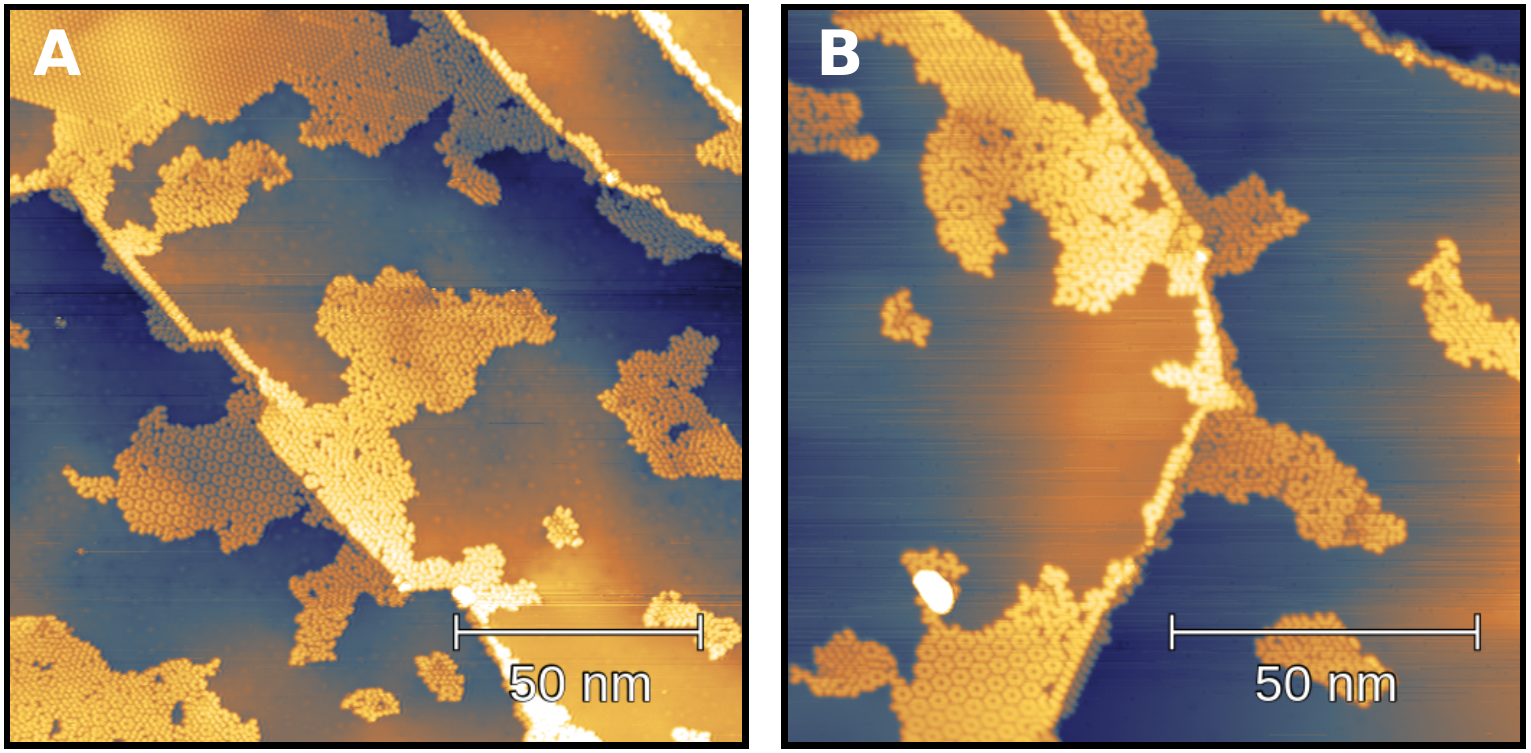}
  \caption{Deposition performed with a filament emission current of \SI{0.1}{\milli\ampere}. The surface contains self-assembled monomers, dimers, hexamers as well as oligomers. Imaging parameters: (A) \SI{-0.85}{\volt}, \SI{120}{\pico \ampere} and (B) \SI{1.13}{\volt}, \SI{32}{\pico \ampere}.}
  \label{fig:0.1mA}
\end{figure}

%\newpage
\textbf{0.1 mA emission:} In this setting, ordered self-assemblies of monomer, dimers and hexamers, as well as mixed assemblies of oligomers were found (Figure \ref{fig:0.1mA}). The monomer assemblies are mostly intact, with some disruption by dimers, etc. (Figure \ref{fig:0.1mA}A). 

%\newpage
\textbf{0.04 mA emission:} The surface is similar to \SI{0.1}{\milli\ampere} emission but with a larger amount of hexamer self-assembly and with a larger portion of the monomer self-assemblies disrupted by rows of dimers or disordered assemblies of a mixture of species.

\begin{figure}[H]
  \includegraphics[width=\columnwidth]{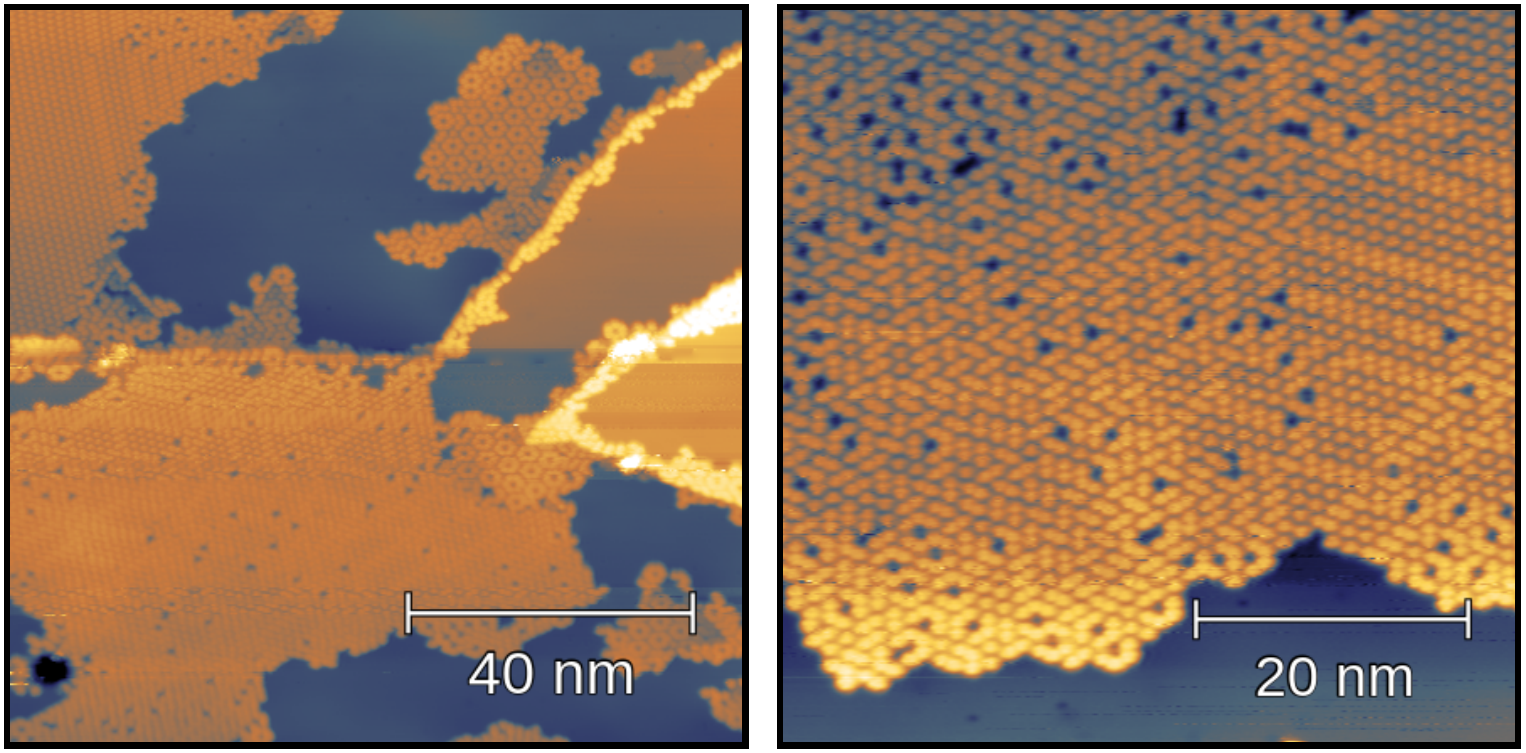}
  \caption{Deposition performed with a filament emission current of \SI{0.04}{\milli\ampere}. The surface looks similar to the results obtained with \SI{0.1}{\milli\ampere}. However, larger areas of the monomer self-assemblies are disrupted with regions containing a mixture of dimers, trimers and other oligomers (B). Imaging parameters: \SI{1.59}{\volt}, \SI{44}{\pico\ampere}.}
  \label{fig:0.04mA}
\end{figure}

The results of the STM data obtained from these three samples is summarized in Table 1 of the main text, and reproduced for convenience in Table \ref{table:StatisticsOfDTPA} below. 
In order to estimate the average number of hydrogen atoms per DTPA molecule in the Table, we assume that every DTPA termination site not bonded to another DTPA molecule is instead bonded with one hydrogen atom. Hence, each DTPA monomer has 3 hydrogen atoms, each dimer has 4 terminal H atoms (two per DTPA molecule), etc. Then, we look at select sample regions with mixed oligomers, as well as the disrupted monomer self-assembly (see, e.g., Figure \ref{fig:0.04mA}B) and obtain an estimate of the average number of DTPA-H bonds per DTPA molecule and an average area density of the DTPA molecules by manually counting the bond order. Finally, using the area densities of the various self-assemblies, we estimate the number of molecules in the scans by manually highlighting the area covered by each of the mentioned categories and estimate the average degree of hydrogenation of the molecules.
\begin{table*}
\centering
\begin{tabular}{llclllc}
Emission & ~~Area & No. molec. & 1-mer & 2-mer & 6-mer & H/DTPA \\
\hline\hline
\SI{0.04}{\milli\ampere} & \SI{153500}{\nano\meter^2}&  \SI{31500} & 13\% & 15\% & 20\% & 1.7\\
\SI{0.10}{} & \SI{126300}{} & \SI{32200} & 36 & \phantom{0}4 & 16 & 2.0\\
\SI{1.00}{} & \SI{226400} & \SI{45500} & 67 & \phantom{0}8 & \phantom{0}- & 2.5\\
\end{tabular}
\caption{Distribution of oligomer species on the Ag(111) substrate for different cracking-filament emission currents. Fractions of molecules in monomer (1-mer), dimer (2-mer), and hexamer (6-mer) SAMs are shown, as well as the total area imaged and the number of molecules. The average number of hydrogen terminations per DTPA molecule (H/DTPA) is calculated considering all molecules, including those in mixed regions of oligomers and partial-COF that do not form well-ordered self-assembly.}
\label{table:StatisticsOfDTPA}
\end{table*}

%\newpage
\section{DTPA Depositions on Au(111)}

\begin{figure}[H]
  \includegraphics[width=\columnwidth]{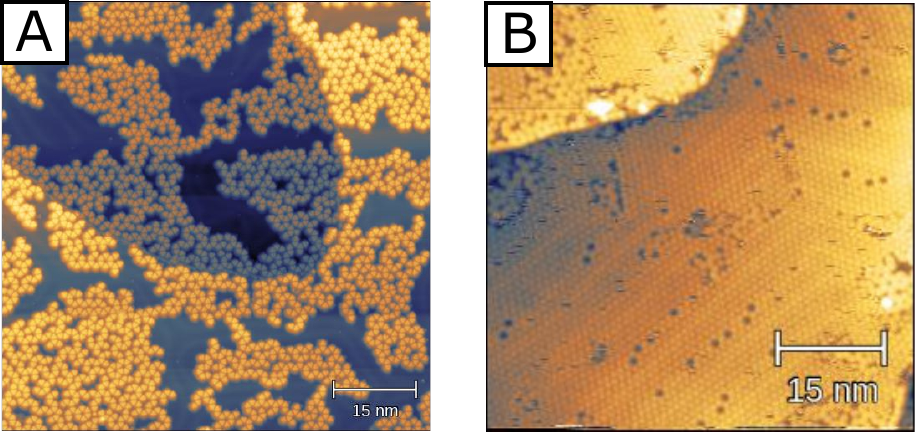}
  \caption{DTPA Molecules deposited on Au(111) substrate in the (A) absence of atomic hydrogen, and (B) in the presence of \SI{2.2E-6}{mbar} of hydrogen in the chamber along with a cracking filament. The depositions were performed at a temperature of \SI[separate-uncertainty=true]{523\pm70}{K}. The large temperature uncertainty is due to experimental issues with sample contact. Imaging parameters: (A) \SI{-1.09}{\volt}, \SI{90}{\pico\ampere} and (B) \SI{1.78}{\volt}, \SI{21}{\pico\ampere}.}
\label{fig:DTPA_Au}
\end{figure}

Depositions prepared for TOF-SIMS experiments were performed on Au(111) substrates rather than the Ag(111) substrate used throughout the text. This was done to avoid potential oxidation of the substrate during transfer to the mass spectrometer. Figure \ref{fig:DTPA_Au} shows the results of the DTPA experiments performed on a Au(111) substrate. Similar to the Ag(111) substrate, the molecules form two dimensional networks when deposited in the absence of a source of atomic hydrogen, while they form a self-assembled structure without bonding when the deposition is performed in the presence of atomic hydrogen.

%%%%%%%%%%%%%%%%%%
%\newpage
\section{ToF-SIMS: Br-terminated and H-terminated DTPA}
\label{sec:TOF}
\begin{figure}[H]
  \includegraphics[width=0.85\columnwidth]{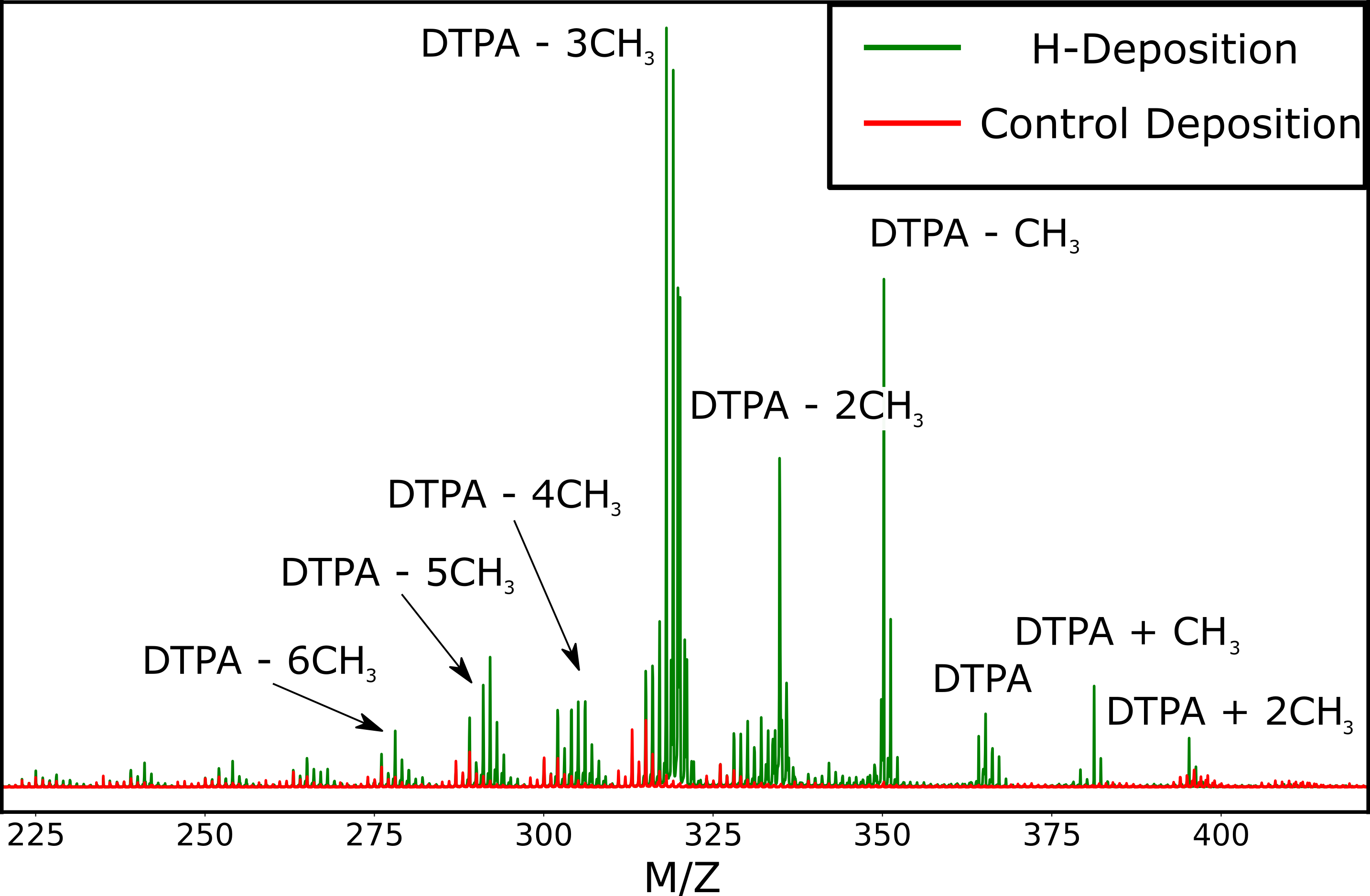}
  \caption{Selection of the ToF-SIMS mass spectrometry data for the non-brominated DTPA mass ranges. The green spectrum is the experimental data for DTPA molecules deposited in a \ce{H2}-rich environment with the cracking filament on. The molecules were deposited onto a Au(111) sample heated to \SI{573}{\kelvin}. The red spectrum is a control deposition of DTPA molecules onto a room temperature Au(111) sample without the presence of additional \ce{H2} and the cracking filament off.}
\label{fig:TOF}
\end{figure}

In this section, we provide additional data for the quantitative evidence of hydrogenation of the molecules. Figure \ref{fig:TOF} shows the mass spectra of the direct deposition of \ch{(Br_3)DTPA} on the Au(111) substrate at room temperature without any atomic hydrogen (red) and deposition on a hot substrate in the presence of atomic hydrogen (green). The peaks of the non-hydrogenated sample are located 3u lower than the corresponding peaks of the hydrogenated sample. The reduction in mass by 3u indicates that the peaks are caused by fragments of \ch{(Br_3)DTPA} that lose the \ch{Br} atoms in the ion generation process, but do not have hydrogen on the three termini.

%%%%%%%%%%%%%%%%%%
%\newpage
\section{OTPA Depositions on Au(111)}
\begin{figure}[H]
  \includegraphics[width=0.75\columnwidth]{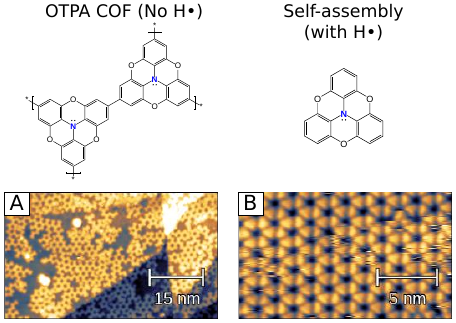}
  \caption{Results of OTPA deposition on Au(111) substrate. (A) \SI{473}{\kelvin}(est.) substrate, in \SI{3E-8}{\milli\bar} of background $\ch{H_2}$ (B) \SI{453}{\kelvin}(est.)  in \SI{2.2E-6}{\milli\bar} of background $\ch{H_2}$ and hot ion gauge filament with an emission current \SI{0.1}{\milli\ampere}. (Faulty thermal contact made the temperature of this substrate unreliable, but the results of subsequent experiments are consistent with the estimated temperatures). Imaging parameters: (A) \SI{0.68}{\volt}, \SI{130}{\pico \ampere} and (B) \SI{1.83}{\volt}, \SI{24}{\pico\ampere}.} 
\label{fig:OTPA_Au}
\end{figure}

Figure \ref{fig:OTPA_Au} shows the results of the deposition of OTPA on the Au(111) surface. These molecules form a two dimensional covalent network when deposited without any sources of atomic hydrogen (A), and form a self-assembled monolayer when deposited in the presence of atomic hydrogen (B). The OTPA molecules image differently in STM as compared to the DTPA molecule; the absence of the nearly vertical methyl bridging groups means that the STM images the OTPA molecules with the brominated sites at the vertices of the triangles(as opposed to the DTPA methyl bridge sites as the vertices). The self-assembled phase we observe is similar to the un-annealed, self-assembled phase of the \ch{(Br_3)OTPA} molecule on Au(111) reported in the literature \cite{marchi_temperature-induced_2019}.

%%%%%%%%%%%%%%%%%%
%\newpage
\section{ToF-SIMS: Deuterium substitution}

\begin{figure}[H]
  \includegraphics[width=.8\columnwidth]{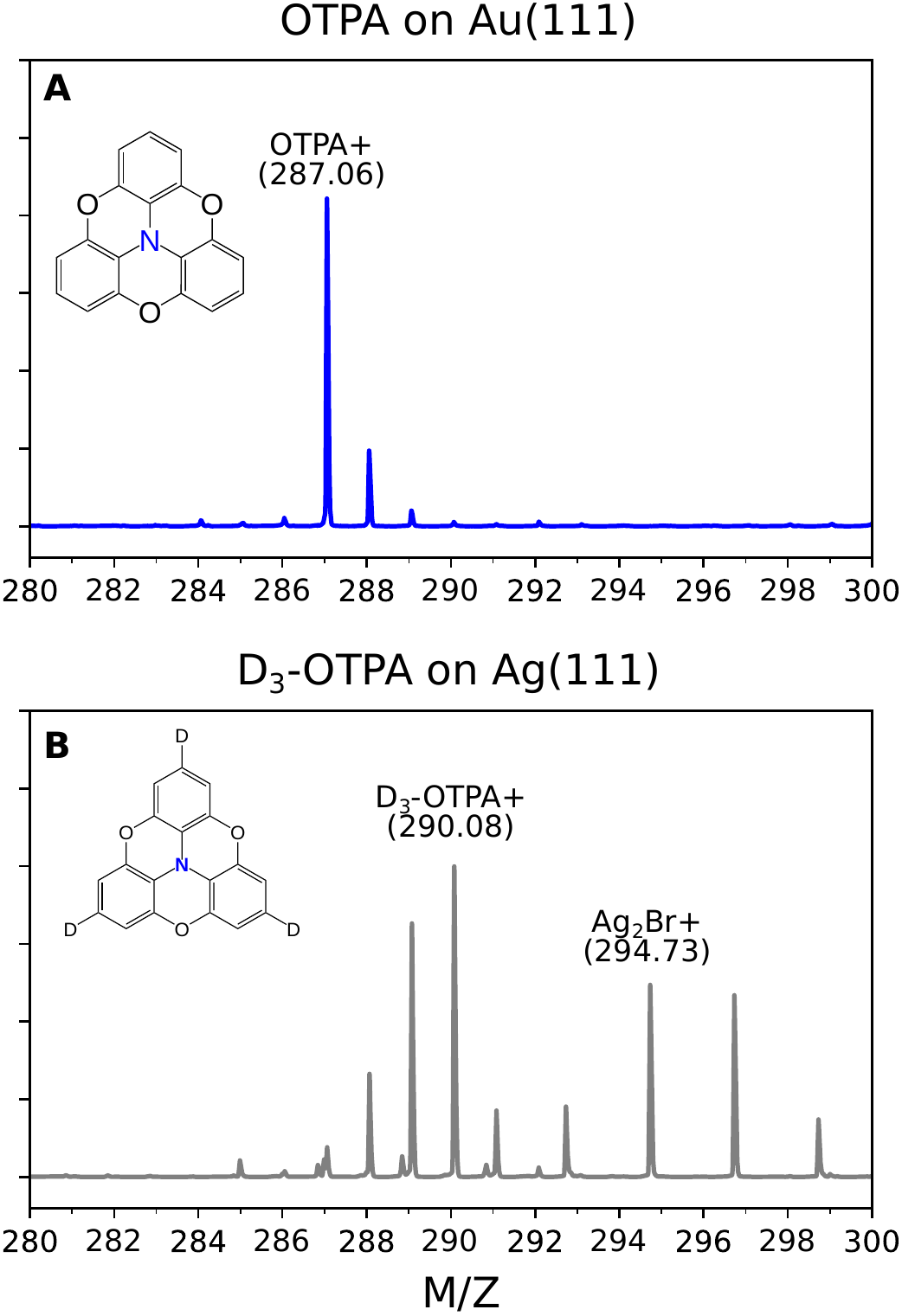}
  \caption{Comparison of the OTPA on Au(111) spectrum (A) and OTPA on Ag(111) spectrum (B). The primary peak in (B) matches the expected mass of the deuterium-terminated OTPA (290.08), providing further evidence of our proposed termination. The cluster of four peaks (292.73, 294.73, 296.73, 298.73) are attributed to \ch{Ag_2Br^+} formed by the surface bromine that is released during dehalogenation of \ch{(Br_3)OTPA}. }
\label{fig:D3-OTPA TOF}
\end{figure}

We also perform a deposition of \ch{(Br_3)OTPA} molecules on a heated Ag(111) substrate in the presence of atomic deuterium (backfilling the chamber with deuterium gas along with a hot ion gauge filament for cracking) and perform ToF-SIMS on the sample, following the same procedure as described in the main paper. We compare the mass spectra of the hydrogenated OTPA on Au(111) sample with the deuterated OTPA on Ag(111) sample (Figure \ref{fig:D3-OTPA TOF}). The primary peak in the deuterated sample is shifted up by 3u (due to the heavier deuterium atoms). In addition to the fully deuterated peak corresponding to \ch{D_3-OTPA}, we also observe mass peaks that consist of partially deuterium and hydrogen substituted species (\ch{HD_2-OTPA} and \ch{H_2D-OTPA}) as well as a peak corresponding to the fully hydrogenated OTPA. We attribute these peaks to the existence of background hydrogen in the chamber in addition to the deuterium that has been introduced (although, we don't exclude the possibility of minor H-D substitution during the 5-min atmospheric exposure during transfer to the TOF-SIMS instrument). Using the areas of the peaks, we estimate that $\approx 68.7 \%$ of the sites are passivated by deuterium, the rest being passivated by hydrogen. The set of four additional peaks at masses 292.7-298.7 are attributed to the \ch{Ag_2Br^+} ions that occur due to bromine adsorbed on the silver surface. 

Hence, this experiment confirms that the source of the hydrogenation of the deposited molecules is the background gas during deposition, as opposed to other sources or atmospheric contamination.

%%%%%%%%%%%%%%%%%%%%%%%%%%
%\newpage
\section{DTPA Commensurate Overlayer}
\label{sec:Overlayer}

\begin{figure}[H]
  \includegraphics[width=\columnwidth]{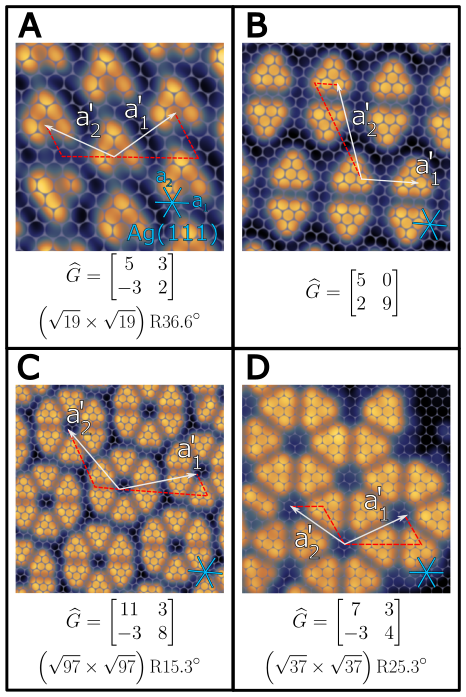}
  \caption{STM topographs of DTPA commensurate molecular overlayers on a synthetic Ag(111) lattice (hexagonal lattice with $a_{Ag(111)} =$ \SI{0.288}{\nano\metre}. Topographs feature overlayer lattice vectors (\(\vec{a}_1'\) and \(\vec{a}_2'\)) with white arrows and their components projected onto the Ag(111) lattice by the red dashed lines. Below the topographs are the matrix and Wood's notation for the commensurate overlayer structures. (A) DTPA monomer SAM. (B) DTPA dimer SAM. (C) DTPA hexamer macrocycle SAM. (D) DTPA COF.}
\label{fig:Commensurate}
\end{figure}

\begin{figure}[H]
  \includegraphics[width=\columnwidth]{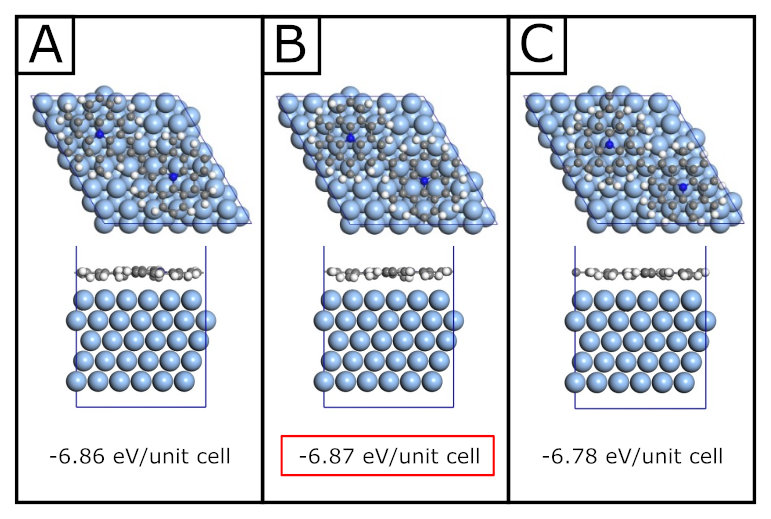}
  \caption{The interface system of \ce{CH2}-bridged DTPA-COF on the Ag substrate was calculated at the DFT-PBE level of theory (see Methodology in the main text). Three adsorption sites, denoted as (A) bridge, (B) hollow, and (C) top sites and characterized by the relative positions of the center-\ce{N} atom in the DTPA COF were considered in the calculations. The relative stability of the three adsorption sites was characterized by comparing their adsorption energies, defined as $E_{ads}= E_{total}-E_{Ag}-E_{COF}$, where $E_{total}$, $E_{Ag}$, and $E_{COF}$ are the energies of COF/Ag complex, the Ag slab, and the COF monolayer, respectively.}
\label{fig:Adsorption}
\end{figure}

In Figure~\ref{fig:Commensurate}, a synthetic Ag(111) lattice is superimposed over STM topographs of SAM structures (to avoid obscuring the molecules, the shaded honeycomb denotes the \emph{interstitial} regions of the Ag(111) surface; Ag atom positions are the transparent circles). Although the Ag lattice was generally not resolved while imaging molecular islands, its orientation was determined by Ag \(\langle1\bar{1}0\rangle\)-aligned step-edges in larger scans over the same region (a minimum of two step directions among \([1\bar{1}0]\), \([10\bar{1}]\), or \([01\bar{1}]\) was used). Without atomic resolution of the Ag lattice, the precise location of a molecule over the substrate was not observed experimentally. The lateral alignment of the lattice shown in Figure~\ref{fig:Commensurate} is chosen to align the DTPA \ce{N}-centers over 3-fold hollow sites, a configuration which is consistent with the lowest-energy configuration of a closely-related \ce{CH2}-bridged TPA COF on the Ag(111) surface, calculated for another purpose (Figure~\ref{fig:Adsorption}).

%%%%%%%%%%%%%%%%%%%%%%%%%%
%\newpage
\section{DFT Adsorption Calculations}
\label{sec:DFTsupp}

%%%%%%%%%%%%%%%%%%%%%%%%%%%%%%%%%
%\newpage
\section{Tight binding calculations}
\begin{figure}[H]
  \includegraphics[width=0.5\columnwidth]{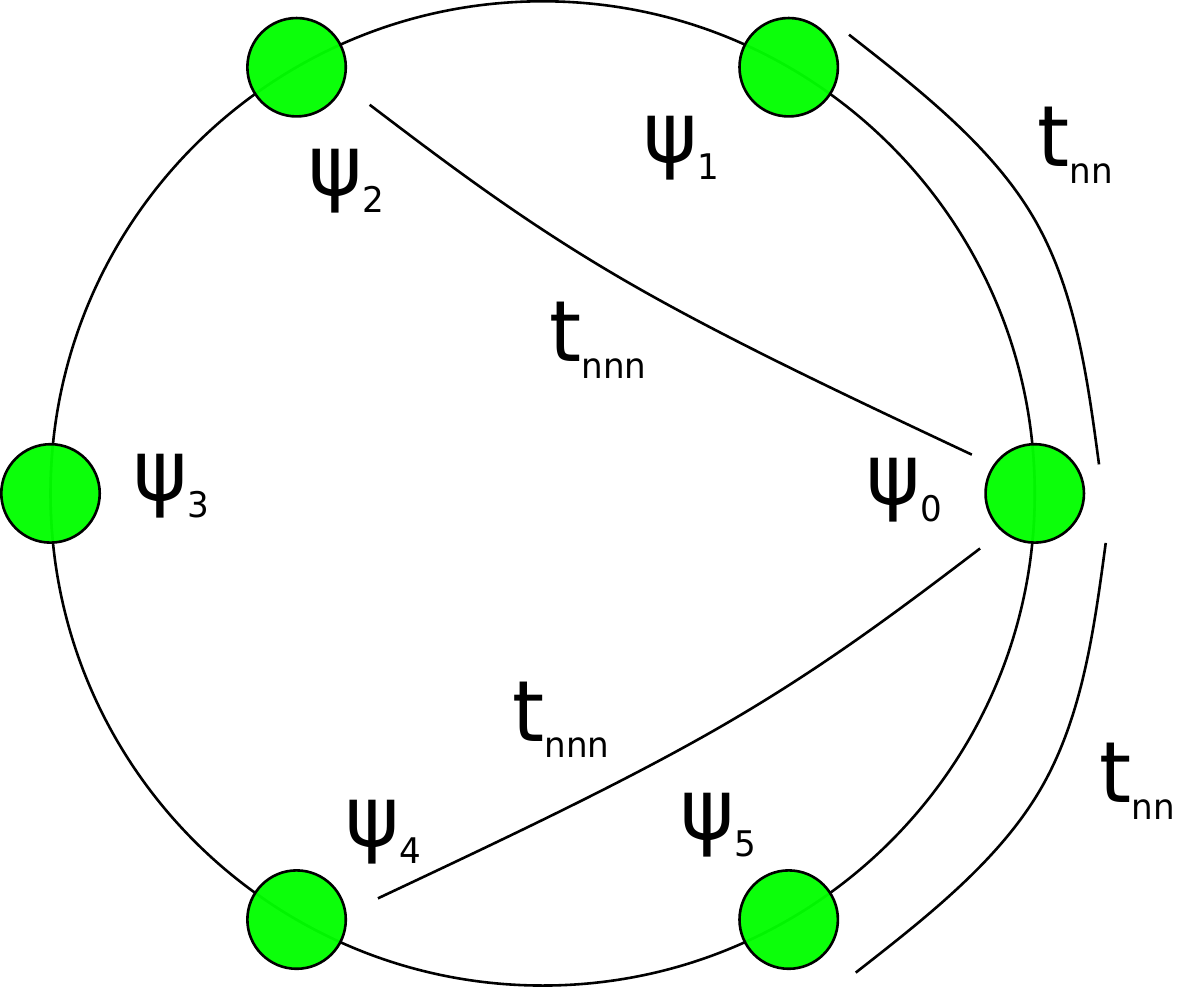}
  \caption{Graphical representation of the tight-binding network used in the case of a 6-membered ring. All (nearest neighbor (nn) and next-nearest neighbor (nnn) are shown for one site ($\psi_1$).}
\label{fig:6_member_ring}
\end{figure}

We consider a simple six-member ring tight binding Hamiltonian with nearest neighbor and next-nearest neighbor terms to model the evolution of the filled state energy levels from monomer to six-membered macrocycle as shown in Figure \ref{fig:6_member_ring}. Since all the sites of the six membered ring are equivalent, we ignore their on-site energies. 

\begin{equation*}
    \mathcal{H} = \sum_{nn} t_{nn} (a_i^\dagger a_j + a_j^\dagger a_i) + \sum_{nnn} t_{nnn} (a_i^\dagger a_j + a_j^\dagger a_i)
\end{equation*}

Now, we consider a ring with $N$ orbital sites.

\begin{equation*}
    t_{nn}(\psi_{n-1}+\psi_{n+1})+t_{nnn}(\psi_{n-2}+\psi_{n+2}) = E\psi_n
\end{equation*}

Here, $\psi_n$ refers to the amplitude of the wavefunction at site $n$ and the indices $n$ wrap around the $N$ member circle. 

This admits solutions of the form $\psi_n = e^{\frac{i 2\pi m n}{N}}$, where $m \in \{0,1,...,N-1\}$, with solutions $E_m = 2t_{nn}cos\left(\frac{2\pi m}{N}\right) + 2t_{nnn}cos\left(\frac{4\pi m}{N}\right)$. For the six-membered ring system which we use, the solutions are singly degenerate levels at $(2t_{nn}+2t_{nnn}, -2t_{nn}+2t_{nnn})$ and doubly degenerate levels at $(t_{nn}-t_{nnn},-t_{nn}-t_{nnn})$.

Comparing the peak locations predicted by this model to the DFT peak locations, we obtain $t_{nn}\simeq \SI{-0.2}{eV}$ and $t_{nnn}\simeq\SI{-0.012}{eV}$.

%%%%%%%%%%%%%%%%%%%%%%%%%%
%\newpage
\section{Freestanding DTPA COF DFT Bandstructure}
\label{sec:DFTsupp2}

\begin{figure}[H]
  \includegraphics[width=\columnwidth]{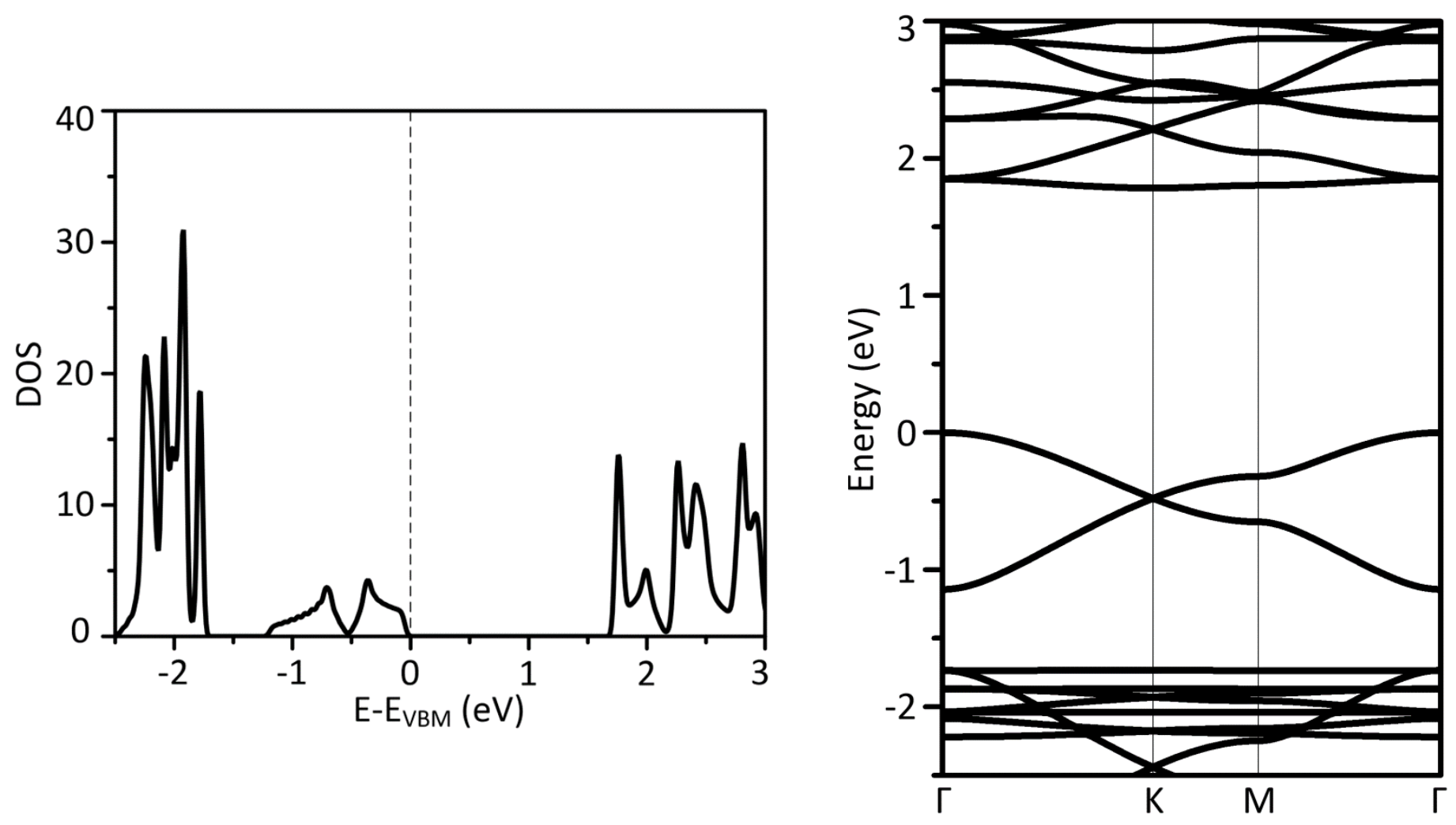}
  \caption{Left: DFT calculated density of states for the freestanding DTPA COF. Right: DFT calculated bandstructure for the freestanding DTPA COF.}
\label{fig:DTPAbands}
\end{figure}

%%%%%%%%%%%%%%%%%%%%%%%%%%%%%%
%\newpage
\section{Spectral Map Analysis}
\label{sec:SpecMap}

\begin{figure}[H]
  \includegraphics[width=\columnwidth]{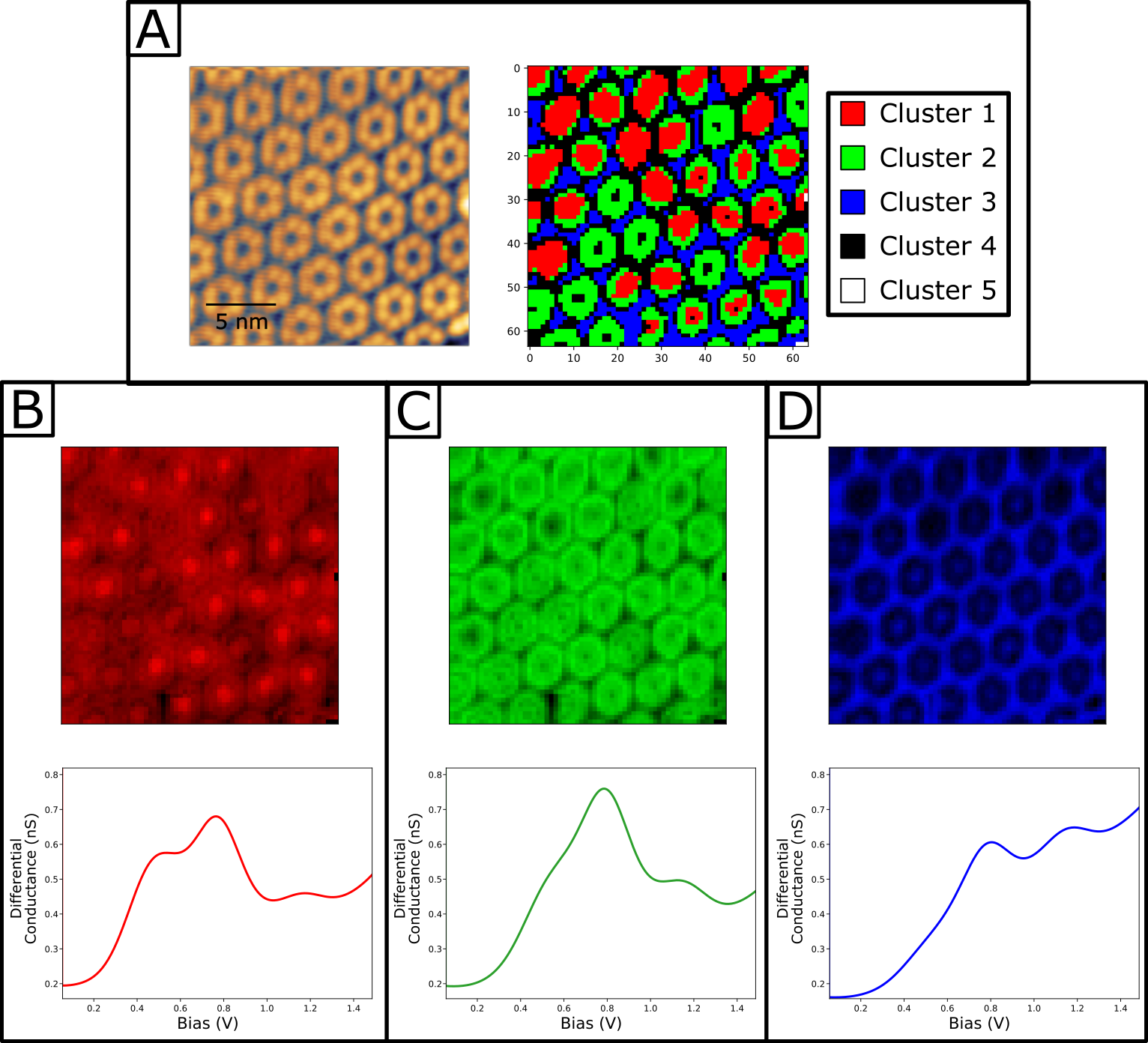}
  \caption{Details on the spectral map analysis using the hexamer SAM data as an example. (A) A reference topograph taken over the hexamer SAM immediately prior to the spectral map acquisition and the sorted results from K-means clustering using five clusters. The K-means results display the 64x64 pixel map where the scanning tunneling spectra are taken and each color designates its cluster. (B-D) Representation of the vector 2-norm in the vector space spanned by the bias voltages with the displayed spectra as the origin. These spectra are the average differential conductance spectra between \SIrange{0.05}{1.5}{\volt} from the selected clusters. The Euclidean distances were used to generate the red (B), green (C), and blue (D) pixel values in middle column of Figure~6.}
\label{fig:KmeanSupp}
\end{figure}

Figure~\ref{fig:KmeanSupp} shows the data analysis steps underlying Figure~6 of the main text.
The spectral map data is first sorted by K-means clustering into 5 clusters. K-means was chosen because it can sort the spectral data without the necessity for user intervention or thresholding.  K-means is performed on the differential conductance spectra in the range from \SIrange{0.05}{1.5}{\volt} which has been normalized by its integral value. Representative spectra are chosen as the origin to calculate a Euclidean distance in the vector space spanned by the bias voltages. These representative spectra are the averaged spectra from cluster 1 (B), cluster 2 (C) and cluster 3 (D). The clusters were chosen based on the spatial portion of the molecular formation under investigation.  The Euclidean distance is the vector 2-norm from the spectrum at a pixel location in the spectral map to the cluster averaged spectrum (cluster center) resulting in three Euclidean distances per pixel location. These values are normalized by the range of values (per spectral origin point) and subtracted by 1 resulting in a value between 0-1 with 1 being closest to the referenced spectrum and 0 furthest away. The sets of these values are shown in the maps presented in (B-D). These maps are converted into the RGB values displayed in Figure 6 with the red pixel value from map (B), green from map (C), and blue from map (D).

%%%%%%%%%%%%%%%%%%%%%%%%%%
%\newpage
\section{Freestanding DTPA DFT Partial Charge Density Maps}
\label{sec:DFTsupp3}
\begin{figure}[H]
  \includegraphics[width=.85\columnwidth]{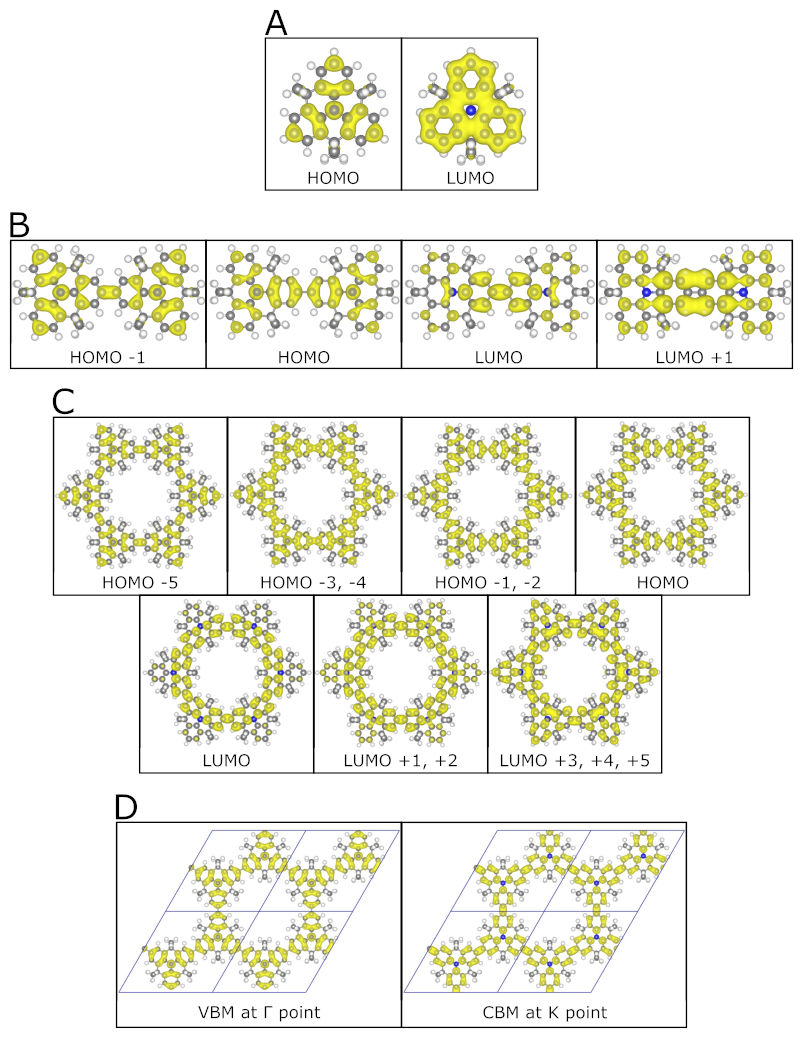}
  \caption{DTPA charge density maps calculated by freestanding DFT for various DTPA structures and molecular orbitals/bands. (A) DTPA monomer. (B) DTPA Dimer. (C) DTPA hexamer macrocycle. (D) DTPA COF.}
\label{fig:DFTcharge}
\end{figure}

%%%%%%%%%%%%%%%%%%%%%%%%%%%%%%%%%%%% 
%\newpage
\section{Electron confined to a circle in two dimensions}

Consider an electron obeying the Schrödinger equation confined to a circle in two dimensions with radius $\rho$\cite{crommie_quantum_1995}. 
\begin{equation*}
    \frac{-\hbar^2}{2m^*}\nabla^2 \psi = E \psi
\end{equation*}

Expanding $\nabla^2$ in polar coordinates, we get:

\begin{equation*}
        \nabla^2 = \frac{\partial^2}{\partial r^2}+ \frac{1}{r}\frac{\partial}{\partial r}+ \frac{1}{r^2} \frac{\partial^2}{\partial \theta^2}
\end{equation*}

Splitting $\psi$ as $R(r)\Phi(\theta)$,

\begin{equation}
    \frac{r^2}{R}\frac{\partial^2R}{\partial r^2}+ \frac{r}{R}  \frac{\partial R}{\partial r}+\frac{2m^*E}{\hbar^2}r^2 = m^2
\end{equation}

where $m \in \mathbb{Z}$ due to the condition on the angular part of the wavefunction ($\Phi(\theta+2m\pi) = \Phi(\theta)$).

Substituting $r'=\sqrt{\frac{2mE}{\hbar^2}}r$, we get the Bessel differential equation:
\begin{equation}
    r'^2\frac{\partial^2R}{\partial r'^2} + r' \frac{\partial R}{\partial r'} + R(r'^2-m^2) = 0
\end{equation}

The solutions of this equation are the Bessel functions. If we constrain the solutions of this equation to be zero at $r=\rho$ and that $R(0)$ isn't a singularity, $R(r')$ at $r=\rho$ must be zero. The lowest energy mode would correspond to the location of the first zero of $J_0(r')$\cite{crommie_quantum_1995}, ie., 
\begin{equation*}
    E = E_0+\frac{\hbar^2}{2m^*} \left(\frac{2.4048}{\rho}\right)^2 
\end{equation*}

Since we are looking at the confinement of surface state electrons with parabolic dispersion, we add the energy term $E_0$, where $E_0$ is the onset of the surface state.

If we take $m^*=0.42m_e$ and an approximate radius of the confinement provided by the hexamers/COF as $\rho=\SI{0.94}{\nm}$ and $E_0=\SI{-50}{meV}$, we get a ground state energy of $E=\SI{0.54}{\eV}$, which approximates the pore resonance observed in the EPWE model.

%%%%%%%%%%%%%%%%%%%%%%%%%%%%%%%%%%%%
\section{Electron Plane Wave Expansion modeling}

The free surface state electrons obey a parabolic energy dispersion given by:
\begin{equation*}
    E_k = E_0 + \frac{\hbar^2 k^2}{2m^*}
\end{equation*}

Here, $E_0$ is the surface state band onset energy (measured to be $\SI{-50}{meV}$ from STS) and $m^*$ is the effective mass of the surface state electrons, $m^*=0.42m_e$\cite{li_electron_1998}.

The adsorbed molecules on the metal surface can be modelled as regions containing a scattering potential that affects the two dimensional surface electron gas \cite{kepcija_quantum_2015,klappenberger_dichotomous_2009,klappenberger_tunable_2011}. The potentials can be used to set up an effective Schrödinger equation. The differential equation can be solved numerically using a variety of methods, such as the Boundary Element Method\cite{kepcija_quantum_2015} or an Electron Plane Wave Expansion (EPWE) method\cite{abd_el-fattah_graphene_2019,kawai_near_2021}. 

In the EPWE method\cite{abd_el-fattah_graphene_2019}, we consider the Schrödinger equation with a periodic potential $V$:
\begin{equation*}
    \left( \frac{-\hbar^2\nabla^2}{2m^\star} + V(\textbf{r})\right) \psi = E \psi
\end{equation*}

The potential $V$ can be expanded in two dimensions using a Fourier series summation as $V(\textbf{r}) = \sum_{\textbf{G}} V_{\textbf{G}} e^{i\textbf{G.r}}$, where the summation is over all the reciprocal lattice vectors $\textbf{G}$ of the periodic potential $V$. We can also expand the Bloch wavefunction $\psi = u(\textbf{r}) e^{i\textbf{k.r}} $ where $u(\textbf{r})$ has the same periodicity as the potential $V(\textbf{r})$ and $\textbf{k}$ is a vector in the first Brillouin zone. Hence, the wavefunctions $\psi$ can also be expanded in terms of a Fourier series of the reciprocal lattice as $\psi(\textbf{r}) = \sum_{\textbf{G'}} u_{\textbf{G'}} e^{i \textbf{(G'+k).r}}$. We insert these terms into the equation. Then, for every reciprocal lattice vector $\textbf{G}_0$, we can write:

\begin{equation*}
    \frac{\hbar^2}{2m^*}(\textbf{G}_0+\textbf{k})^2 u_{\textbf{G}_0} + \sum_{\textbf{G}} V_{\textbf{G}_0-\textbf{G}} u_{\textbf{G}} = E u_{\textbf{G}_0}
\end{equation*}

While this equation is for all reciprocal lattice vectors $\textbf{G}_0$, for the purpose of the numerical calculation, we restrict the set of lattice vectors to within a circle in the reciprocal space with a radius sufficiently large (with sufficiently many Fourier components) to describe the potential accurately and obtain convergence. Hence, this equation can be written as an eigenvalue equation $\mathcal{H}U=EU$, where the matrix $\mathcal{H}_{ij} = V_{\textbf{G}_i-\textbf{G}_j}+\frac{\hbar^2}{2m^*}(\textbf{G}_i+\textbf{k})^2\delta_{ij}$. This eigenvalue equation can be solved numerically for $\textbf{k}$ to obtain the wavefunction energies and densities.

To approximate the DTPA molecules, we constructed a model potential of Gaussian potentials, $V(\textbf{r})= W e^{\frac{-\textbf{r}^2}{2\sigma^2}}$, at every atomic site of the DTPA $\pi$-skeleton. For these computations, we choose $W=\SI{0.5}{\eV}$ and $\sigma = \SI{0.7}{\angstrom}$.

\begin{figure}[H]
  \includegraphics[width=\columnwidth]{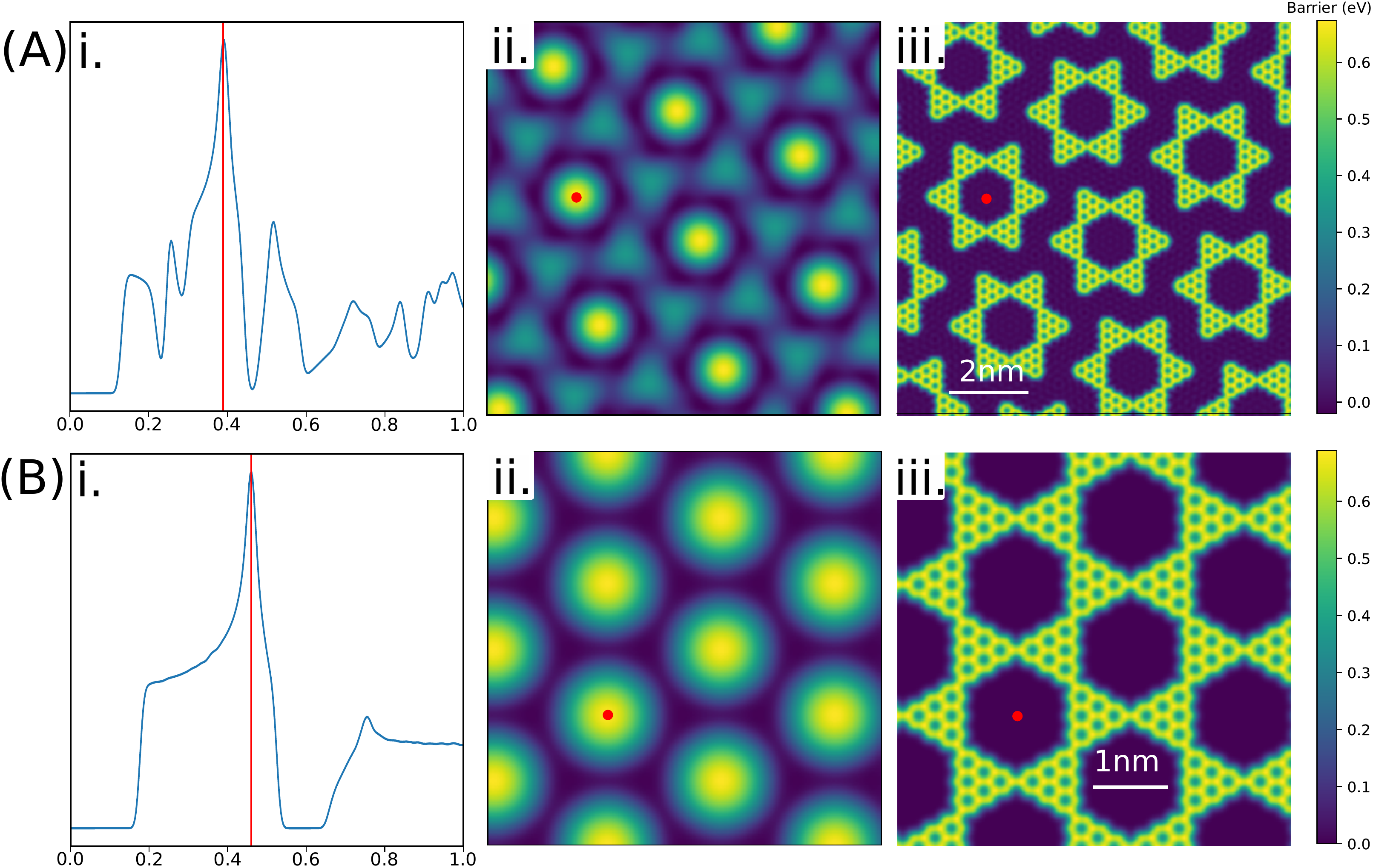}
  \caption{The calculated (i) LDOS at pores, (ii) LDOS intensity map for the pore resonance peak and (iii) Potential used for EPWE calculations for (A) Hexamer self-assembly and (B) COF respectively. The red line in (i) marks the energy of the calculated LDOS map. The red point in (ii),(iii) shows the location (inside the pores) of the spectrum shown in (i). The spectra are broadened in energy with a gaussian kernel ($\sigma=\SI{8.5}{meV}$).}
\label{fig:LDOS_maps}
\end{figure}

Using these model molecular potentials, we can calculate the LDOS of the surface state electrons with confining hexamer as well as COF potentials, as shown in Figure \ref{fig:LDOS_maps}. For this calculation, we consider all the reciprocal lattice vectors $\textbf{G}$ that are inside a circle of radius $15|\textbf{k}_0|$, where $|\textbf{k}_0|$ is the magnitude of the reciprocal lattice vectors of the lattices being considered (COF and hexamer) in order to accurately capture the potential. We truncate the higher order Fourier components since they do not contribute significantly to the low energy confinement LDOS. The LDOS is then calculated over one rhombus shaped lattice tile enclosed by the lattice vectors $\textbf{a}_1$ and $\textbf{a}_2$ in a $50\times 50$ grid. The rhombus tile is then repeated to tile the image shown in Figure~\ref{fig:LDOS_maps}, and the spectra at points of interest are extracted by linear interpolation within the tiled set of spectra. The calculated LDOS spectra are then smoothed in energy using a gaussian filter with a standard deviation of $\SI{8.5}{meV}$ to reflect the experimental broadening.

The pore spectrum of the hexamer shows a peak with a maximum at \SI{0.39}{eV}, while the pore spectrum of the COF shows a maximum at \SI{0.46}{eV}. The LDOS maps plotted at these voltages show that the intensity is concentrated at the pores, indicating that this peak corresponds to a pore resonance. The LUMO states also have higher density towards the pores (Figure \ref{fig:DFTcharge}), which contributes to the higher density of states towards the pore regions. This matches with the experimental observation of increased differential conductance at the pore centers in the case of both hexamer and COF spectra in the region of $\sim\SI{0.5}{eV}$.

\bibliography{SI-references}